\documentclass[acmsmall]{acmart}
\usepackage{multirow}
\usepackage{multicol}
\usepackage{caption}
\usepackage{pifont}
\usepackage{longtable}
\usepackage{diagbox}
\usepackage{amsmath}
\usepackage{tikz}
\usepackage{colortbl}
\usepackage{graphicx}
\usepackage{tikz}
\usepackage{pgfplots}
\usepackage{pgf-pie}
\usepackage{subcaption}
\usepackage{booktabs}
\usepackage{hyperref}
\usepackage[normalem]{ulem}
\useunder{\uline}{\ul}{}

\pgfplotsset{compat=1.18}
\renewcommand\footnotetextcopyrightpermission[1]{} 
\AtBeginDocument{%
  }
\newcommand{\mycheckmark}{\textcolor{black}{\ding{51}}}

\acmJournal{JACM}

\begin{document}

\title{A Survey on Bundle Recommendation: Methods, Applications, and Challenges}

\author{Meng Sun}
\email{sunmeng@whut.edu.cn}
\author{Lin Li}
\email{cathylilin@whut.edu.cn}
\affiliation{%
  \institution{Wuhan University of Technology}
  \city{Wuhan}
  \country{China}
}

\author{Ming Li}
\email{liming7677@whut.edu.cn}
\affiliation{%
  \institution{Wuhan University of Technology}
  \city{Wuhan}
  \country{China}
}
\affiliation{%
  \institution{York University}
  \city{Toronto}
  \country{Canada}
}

\author{Xiaohui Tao}
\email{Xiaohui.Tao@unisq.edu.au}
\affiliation{%
  \institution{University of Southern Queensland}
  \city{Toowoomba}
  \country{Australia}
}

\author{Dong Zhang}
\email{zhangdong23@whut.edu.cn}
 \author{Qing Xie}
\email{felixxq@whut.edu.cn}
\affiliation{%
 \institution{Wuhan University of Technology}
 \city{Wuhan}
 \country{China}
 }

\author{Peipei Wang}
\email{ppwang@sdas.org}
\affiliation{%
\institution{Shandong Computer Science Center (National Supercomputer Center in Jinan), Qilu University of Technology (Shandong Academy of Sciences)}
 \city{Jinan}
 \country{China}
}

 \author{Jimmy Xiangji Huang}
 \email{jhuang@yorku.ca}
 \affiliation{%
 \institution{York University}
 \city{Toronto}
 \country{Canada}
 }

\renewcommand{\shortauthors}{M. Sun et al.}
\begin{abstract}
In recent years, bundle recommendation systems have gained significant attention in both academia and industry due to their ability to enhance user experience and increase sales by recommending a set of items as a bundle rather than individual items. This survey provides a comprehensive review on bundle recommendation, beginning by a taxonomy for exploring product bundling. We classify it into two categories based on bundling strategy from various application domains, i.e., discriminative and generative bundle recommendation. Then we formulate the corresponding tasks of the two categories and systematically review their methods: 1) representation learning from bundle and item levels and interaction modeling for discriminative bundle recommendation; 2) representation learning from item level and bundle generation for generative bundle recommendation. Subsequently, we survey the resources of bundle recommendation including datasets and evaluation metrics, and conduct reproducibility experiments on mainstream models. Lastly, we discuss the main challenges and highlight the promising future directions in the field of bundle recommendation, aiming to serve as a useful resource for researchers and practitioners. Our code and datasets are publicly available at \href{https://github.com/WUT-IDEA/bundle-recommendation-survey}{https://github.com/WUT-IDEA/bundle-recommendation-survey}.
\end{abstract}




\maketitle
\section{Introduction}
Recommendation systems (RS) have become essential tools for alleviating information overload. Over the years, item recommendation systems have been widely applied to recommend item to user or users across various domains, achieving significant success \cite{gomez2016netflix, das2007google, briand2021semi, cheuque2019recommender, gao2023cross}. For example, in news recommendation, users' reading history and interests are analyzed to recommend relevant articles, thereby increasing user engagement and satisfaction \cite{das2007google, liu2010personalized, phelan2009using, okura2017embedding}. In e-commerce, recommendation systems have become a cornerstone for boosting sales conversion rates by recommending products related to users' browsing or purchase history, as evidenced by the sophisticated algorithms used by platforms like Amazon \cite{mcauley2015inferring}. Similarly, video streaming platforms like YouTube employ deep learning techniques to ensure users can discover videos or series they are interested in, significantly increasing watch time and user retention \cite{covington2016deep}. Netflix’s recommendation system stands out as a prime example of leveraging complex algorithms and data analysis to provide highly personalized viewing recommendations, which has significantly enhanced user experience and subscription retention \cite{gomez2016netflix}. Music streaming services like Spotify use collaborative filtering \cite{schafer2007collaborative, sarwar2001item, breese2013empirical} and content-based methods \cite{lops2011content, pazzani2007content, lu2015content} to recommend individual tracks tailored to users' preferences, improving user satisfaction and engagement \cite{koren2009matrix, van2013deep}. Additionally, social media platforms such as Facebook and Twitter utilize recommendation systems to deliver personalized content to users, thereby boosting user interaction and time spent on the platform \cite{guy2010social, kazai2016personalised}. Even in the travel and hospitality industry, recommendation systems help users find ideal destinations, accommodations, and activities based on their past behaviors and preferences \cite{linden2003amazon}.

Existing recommendation systems, however, still face challenges in addressing the diverse and personalized needs of users, despite significant progress in various application domains. Particularly, traditional single-item recommendation may not meet users' comprehensive needs when users need to make choices across multiple categories or items. For example, a music lover may prefer recommendations that include a curated album from a favorite artist or a list of music songs within a specific genre that matches his or her mood. As a result, \textbf{bundle recommendation system (Bundle RS)} has emerged \cite{BGCN, BundleGT, MIDGN, CrossCBR, MultiCBR, CoHEAT}, which extends the traditional item recommendation to a set of items that are likely to interest the user when considered together. These approaches not only enhance user satisfaction by providing more holistic recommendations but also have the potential to increase the economic value for providers through up-selling and cross-selling opportunities.

\paragraph{\textbf{Ubiquity of Bundle RS}}
We conduct a high-level overview of recommendation systems, including item recommendation, group recommendation, bundle recommendation, and complex set recommendation. As shown in Table \ref{different tasks in RS}, item recommendation is to recommend a single item to an individual user based on their preferences, known as 1-to-1 recommendation. Group recommendation is to recommend an item to a group of users, referred to as 1-to-N recommendation \cite{wang2021socially, wang2022social, wang2021bert, yang2024multiview, zhao2024DHMAE}. Complex set recommendation is more complicated, recommending a set of items to a group of users, called N-to-N recommendation, such as recommending a travel package to a tour group. Unlike the above three tasks, the focus of this survey is on bundle recommendation, which recommends sets or bundles of items to a user, termed N-to-1 recommendation. This task has been emergingly applied in various sectors such as e-commerce, entertainment, and tourism due to its ability to satisfy and enhance user experience and drive business value. Consequently, bundle recommendation has become a critical task in modern recommendation systems.
\begin{table}[h]
\scriptsize
\vspace{-0.3cm}
\captionsetup{justification=centering}
\caption{Recommendation System Tasks}
\vspace{-0.3cm}
\label{different tasks in RS}
\centering
\resizebox{0.75\textwidth}{!}{
\begin{tabular}{ccc}
\toprule
\textbf{Task}                              & \textbf{Output}                                       & \textbf{Alias}  \\ \hline
Item  Recommendation& A single item to an individual user      & 1-to-1          \\
Group Recommendation                       & A single item to a group of users        & 1-to-N          \\
\textbf{Bundle Recommendation}             & \textbf{A set of items to a single user} & \textbf{N-to-1} \\
Complex Set Recommendation                 & A set of items to a group of users       & N-to-N          \\ \bottomrule
\end{tabular}
}
\vspace{-0.5cm}
\end{table}

\begin{figure}[htbp]
    \centering
    \includegraphics[width=\linewidth]{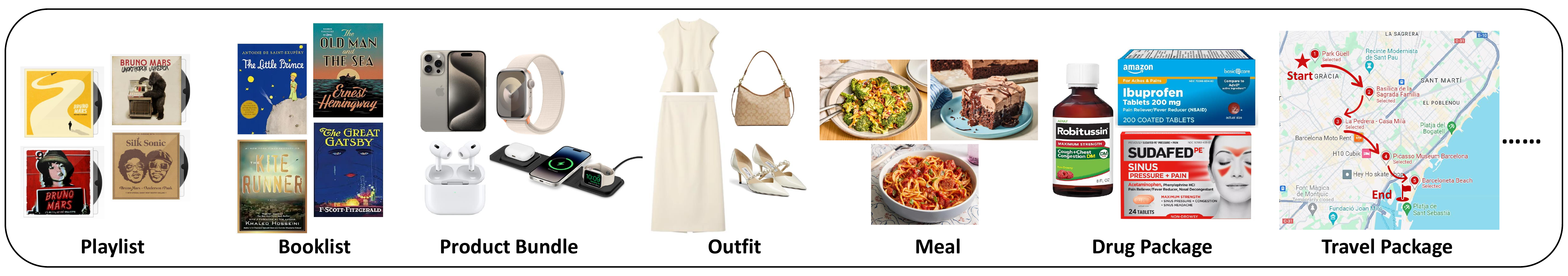}
    \vspace{-0.5cm}
    \caption{Different Application Domains of Bundles}
    \label{fig:different application domains of bundle}
    \vspace{-0.5cm}
\end{figure}

For illustration, several example bundles in the domain of Product, Clothing, Food, Entertainment, Health etc. are depicted in Figure \ref{fig:different application domains of bundle}. In e-commerce, Bundle RS assists retailers in offering product bundles tailored to specific customer needs and preferences, thereby enhancing the shopping experience and boosting sales \cite{liu2017modeling, LIRE, wan2018representing, zhu2014bundle}. For instance, when an electronics enthusiast is looking to buy an iPhone, Bundle RS might recommend complementary items such as AirPods, Apple Watch, and 3-in-1 charger. The electronic product bundles are often sold at a discount, further encouraging the purchase. Similarly, in fashion outfit recommendation \cite{liu2012hi, han2017learning, FPITF, mcauley2015image, POG, HFGN, FHN}, the fourth example of Figure \ref{fig:different application domains of bundle} shows an outfit, including blouse, skirt, heels and a handbag. Suppose that a customer is shopping online with an intent, e.g., purchasing fashion clothing for a party, and finds this well-matched outfit appealing. In this case, the customer may purchase this outfit.

In the entertainment industry, streaming services like Netflix, Spotify, and NetEase use Bundle RS to recommend sets of movies, TV shows, or songs, providing users with a more comprehensive and enjoyable viewing or listening experience \cite{gomez2016netflix, EFM, BGCN, he2020consistency, yang2019music}. Similarly, platforms like Youshu and Goodreads provide lists of books created by readers, leveraging Bundle RS to offer curated book recommendations \cite{DAM, AttList, LIRE}. These bundles enhance the reading experience by recommending books that complement each other, whether based on themes, genres, or reader preferences. Meal recommendation \cite{agapito2017dietos, elsweiler2015towards, MealRec, CMRec, li2023user} is another application of bundle recommendation in food scenarios. For instance, a user looking for a dinner idea might receive a recommendation for a three-course meal featuring a broccoli salad, pasta, and an ice cream cake for dessert. 
This holistic approach to meal planning simplifies decision-making and can inspire users to try new recipes and ingredients. In travel package recommendation \cite{herzog2019integrating, lim2015personalized, lim2018personalized}, suppose a family is planning a vacation to an unfamiliar city; Bundle RS will recommend a travel package with an ordered set of Points of Interest (POIs) that align with their preferences and trip constraints. This makes trip planning more convenient and ensures a memorable, well-rounded experience. Since bundles are ubiquitous across application domains, Bundle RS is widely used in e-commerce, entertainment, and travel.

\paragraph{\textbf{Increasing Importance of Bundle RS}}
Bundle RS has become crucial in enhancing user experiences and boosting business success. In a world saturated with choices, users often feel overwhelmed by decisions. By offering personalized bundles, such as book lists, song lists, meals, or travel package, Bundle RS simplifies decision-making and provides cohesive, enjoyable experiences for users. This task not only increases user satisfaction and loyalty but also drives higher sales, as seen in e-commerce where tailored product bundles are used to meet specific user needs and preferences.

Furthermore, Bundle RS can create highly targeted bundles that evolve over time, aligning with changing preferences and market trends by analyzing user historical data. Research community has also extensively explored Bundle RS, contributing to its development and refinement in both methodology and practice. As technology and user expectations continue to evolve, Bundle RS remains crucial in delivering comprehensive, personalized recommendations.

\paragraph{\textbf{Motivation of this Survey}}
With the growing demand and development of bundle recommendation across various domains, numerous studies have been conducted to improve its performance. Initially, Zhu et al. \cite{zhu2014bundle} proposed the bundle recommendation problem and tried to resolve it by minimizing the cost or maximize the revenue of a bundle \cite{zhu2014bundle, beladev2016recommender}. Later, association rule mining was utilized in \cite{guo2006collaborative, fang2018customized} for bundle generation and recommendation. When a certain number of user-bundle interactions were available, an intuitive solution was to treat the bundle as a single \textquotesingle item\textquotesingle\ and apply traditional collaborative filtering methods. Prior work \cite{FPMC} just ignored the affiliated items of the bundle and just used an id to represent a bundle. Recognizing the importance of affiliated items, some works \cite{EFM, DAM} tried to capture additional user-item interaction and bundle-item affiliation relations. Over time, the bundle view and item view were gradually introduced in bundle recommendation research \cite{BGCN, CrossCBR} etc. This shift led to an increasing amount of research focused on these two views within bundle recommendation \cite{HBGCN, HyperMBR, DGMAE} etc. Considering the increasing attention, emerging challenges, and urgent need for innovative breakthroughs on bundle recommendation, we realize it is the right time to present a survey of this area, providing a systematic review of the methods, applications, and challenges in Bundle RS.

\paragraph{\textbf{Uniqueness of this Work}}
Given the significance and popularity of recommendation research, there are more and more surveys reviewing RS from different perspectives. These include explainable recommendation \cite{explainableRec}, knowledge-based recommendation \cite{Tarus2018}, recommendation methods based on deep learning and reinforcement learning \cite{chen2023deep}, accuracy-oriented recommendation modeling \cite{Accuracy-Oriented}, bias and debias in RS \cite{chen2023bias}, fairness in RS \cite{deldjoo2024fairness, wang2023survey}, and so on. A survey on bundle recommendation by \cite{bundleRStechniques} published in 2020 primarily reviews the main model methods of bundle RS, including integer programming, association analysis, and both traditional and deep learning-based recommendation techniques. These methods have provided effective solutions in specific application scenarios. Along with the development of new techniques such as graph learning \cite{BGCN, CMRec, BundleGT}, contrastive learning \cite{MIDGN, MultiCBR, CrossCBR}, and knowledge distillation \cite{DGMAE}, the research on bundle recommendation has made further progress in the past few years. Therefore, different from existing surveys, our survey offers a more comprehensive and up-to-date review on bundle recommendation, covering the datasets across various bundle recommendation scenarios, both discriminative and generative bundle recommendation tasks, representation learning approaches and strategies, as well as interaction prediction and bundle generation methods. This survey benefits for researchers and practitioners who want to keep up with the state-of-the-art research in the field of bundle recommendation. By doing so, we deliver the following contributions in this survey:
\begin{itemize}
\vspace{-0.1cm}
    \item Based on demands of different application scenarios, i.e., selecting existing bundles or generating new bundles, we introduce a taxonomy for bundle recommendation tasks: discriminative bundle recommendation and generative bundle recommendation. By introducing this clear distinction, we provide two general frameworks tailored to each category and help to identify the appropriate approach for a given task.
    \item To categorize and summarize the existing technologies on discriminative bundle recommendation, we present a comprehensive review that covers representation learning approaches and strategies from bundle and item levels, as well as interaction modeling methods. This detailed review guides researchers to explore more effective solutions in this filed.
    \item To offer a thorough understanding of current generative bundle recommendation, we conclude representation learning methods from item level, and bundle generation approaches. This survey helps researchers identify key methodologies and inspires the development of more effective generative models for bundle recommendation.
    \item A summary of resources for bundle recommendation is provided, including datasets and evaluation metrics. Moreover, we conduct reproducibility experiments on discriminative and generative bundle recommendation models.
    \item We discuss the main challenges and future directions in bundle recommendation from four perspectives. By highlighting these challenges, we aim to guide future research towards overcoming these obstacles and advancing the field.
    \vspace{-0.2cm}
\end{itemize}

\begin{figure}[htbp]
    \centering
    \begin{subfigure}[b]{0.49\textwidth}
        \centering
        \includegraphics[width=\textwidth]{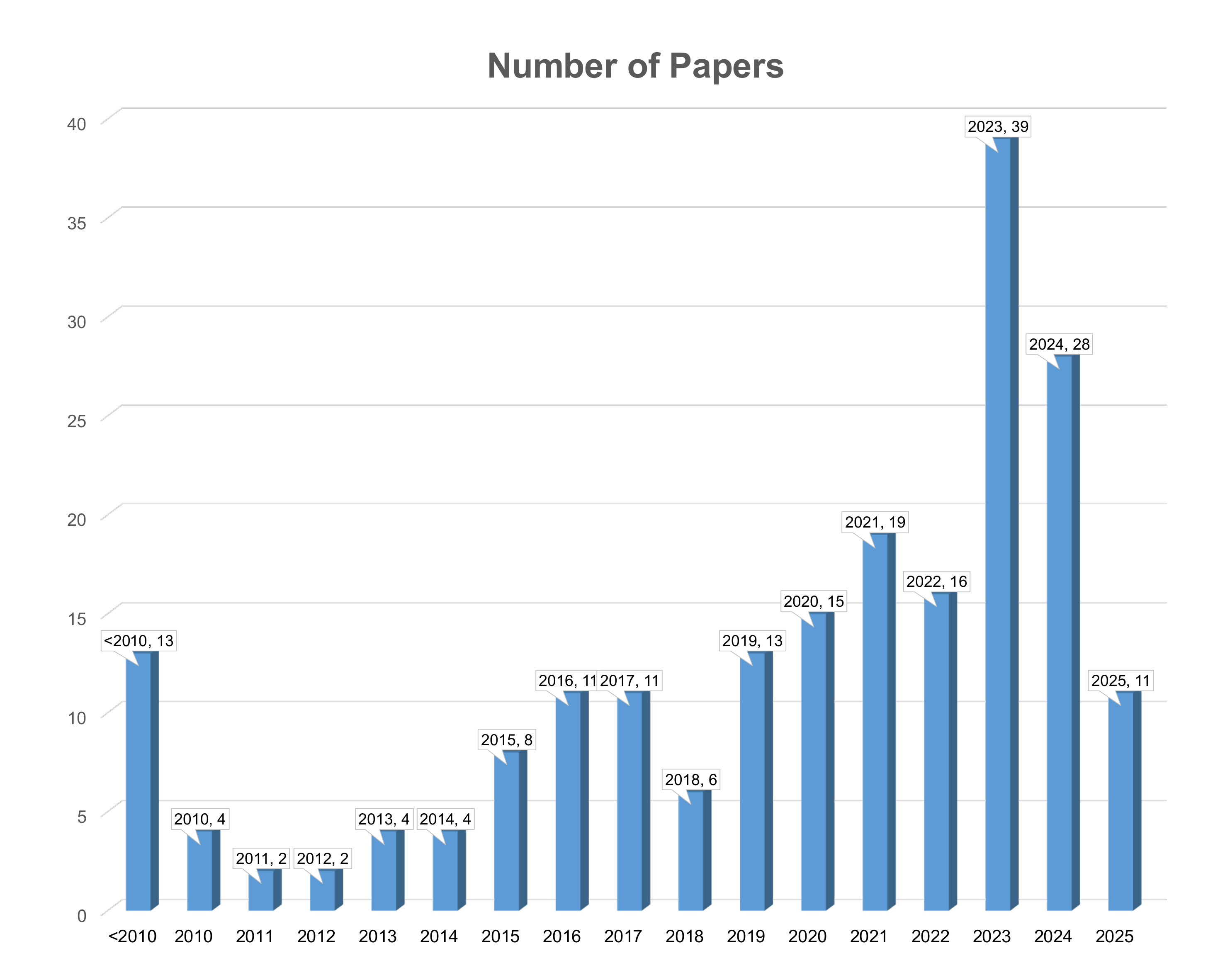}
        \caption{The statistics of papers with the publication year}
        \label{The statistics of publications with publication year}
    \end{subfigure}
    \hfill
    \begin{subfigure}[b]{0.5\textwidth}
        \centering
        \includegraphics[width=\textwidth]{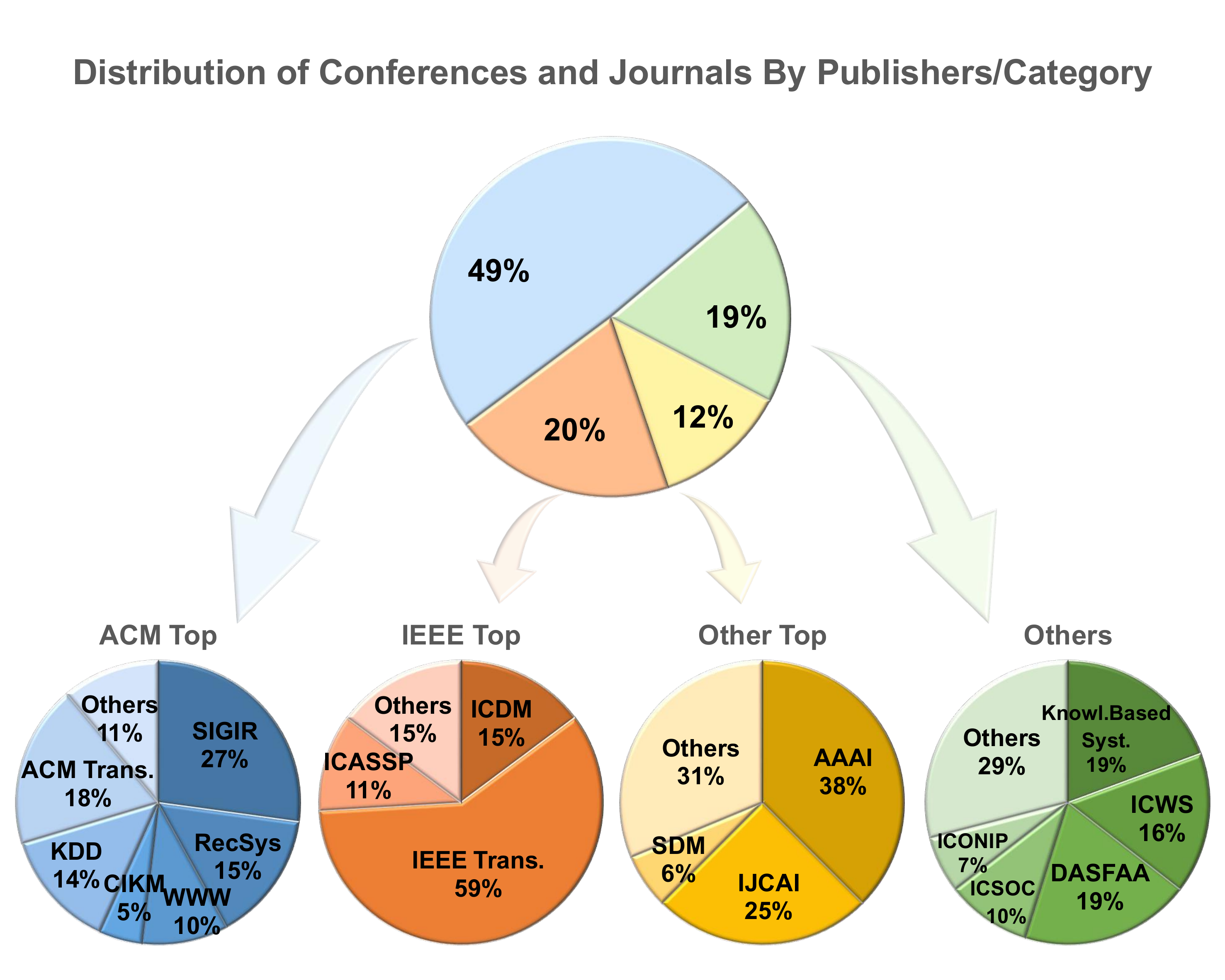}
        \caption{Distribution of papers}
        \label{Distribution of Conferences and Journals By Publisher/Category}
    \end{subfigure}
    \caption{The statistics and distribution of papers related to bundle recommendation}
    \vspace{-0.3cm}
\end{figure}

\paragraph{\textbf{Method of Paper Collection}}
Our survey focuses on reviewing bundle recommendation from the perspective of recommending existing bundles (discriminative) or new bundles (generative), so we retrieved top conferences such as \textit{SIGIR, KDD, WWW, RecSys, WSDM, }etc., and top journals such as \textit{IEEE Trans. Knowl. Data Eng., ACM Trans. Inf. Syst., ACM Trans. Knowl. Discov. Data} and so on. Leveraging scholarly databases like dblp and Google Scholar, we systematically conducted searches employing specific keywords like "\textit{bundle recommend"}, "\textit{package recommend"}, "\textit{list recommend" }, \textit{"next basket recommend"} to search related work. In order to make the retrieved papers more comprehensive, we also used keywords that related to bundle recommendation in different domains, such as \textit{"meal recommend"}, \textit{"outfit recommend",} to get more papers. 
Then, based on the above retrieved papers related to bundle recommendation, we illustrate the statistics of them according to the publication time and publisher, as shown in Figure \ref{The statistics of publications with publication year} and Figure \ref{Distribution of Conferences and Journals By Publisher/Category}, respectively.

\paragraph{\textbf{Survey Organization}}
This survey is structured into seven sections and is organized as follows: Section \ref{section2:preliminaries} introduces the definitions and taxonomy of bundle recommendation. Section \ref{section3:discriminative} discusses the discriminative bundle recommendation task, reviews the representation learning techniques and strategies from bundle and item levels, and summarizes several approaches of interaction modeling. Section \ref{section4:generative} explores the generative bundle recommendation task, and reviews representation learning from item level and bundle generation methods. Section \ref{section5:applications of BR} summarizes the resources of bundle recommendation. Section \ref{section6:challenges and trends} discusses the challenges and trends. And Section \ref{section7:conclusion} concludes the survey.

\section{Definitions and Taxonomy}
\label{section2:preliminaries}
In this section, we introduce the lifecycle of bundle recommendation, providing an overview of its key stages. Based on it, we define the necessary notations, formulate the problem and task of bundle recommendation, and summarize the abbreviations used in Table \ref{Nomenclature} for clarity. Finally, we categorize bundle recommendation into two distinct types based on bundle composition strategy.
\begin{table}[htbp]
\captionsetup{justification=centering}
\caption{Full Names and Abbreviations}
\label{Nomenclature}
\centering
\resizebox{0.85\textwidth}{!}{
\begin{tabular}{ll|ll}
\toprule
\textbf{Full Name}                   & \textbf{Abbreviation} & \textbf{Full Name}          & \textbf{Abbreviation} \\ \hline
Recommender System                   & RS                    & User-Bundle Interaction     & U-B                   \\
Bundle Recommendation System         & Bundle RS             & User-Item Interaction       & U-I                   \\
Discriminative Bundle Recommendation & Discriminative BR     & Bundle-Item Interaction     & B-I                   \\
Generative Bundle Recommendation     & Generative BR         & Knowledge Distillation      & KD                    \\
Graph Neural Network                 & GNN                   & Point of Interest           & POI                   \\
Convolutional Neural Network         & CNN                   & Latent Dirichlet Allocation & LDA                   \\
Recurrent Neural Network             & RNN                   & Term Transition             & TT                    \\
Generative Adversarial Network       & GAN                   & Determinantal Point Process & DPP                   \\
Long Short-Term Memory               & LSTM                  & Natural Language Processing & NLP                   \\
Collaborative Filtering              & CF                    & Large Language Model        & LLM                   \\ \bottomrule
\end{tabular}
}
\vspace{-0.5cm}
\end{table}

\subsection{Lifecycle of Bundle Recommendation}
The efficacy of a bundle recommender system relies on its ability to dynamically evolve with user preferences and behavior. From a macro perspective, the lifecycle of the bundle recommendation is interconnected through three principal components: \textbf{User}, \textbf{Data}, and \textbf{Model}.
\begin{itemize}
    \item \textbf{User.} This component represents the end-users of the bundle RS, who can be customers, readers, listeners, or any individuals engaging with the platforms such as Spotify, Netflix, etc.
    \item \textbf{Data.} It consists of three main types. The first type is the interaction data between the user and single items or bundles, which can be implicit or explicit feedback. The second type is the affiliation data, which details the relationships between bundles and their constituent items. The third type is side information including descriptions of entities, etc.
    \item \textbf{Model.} This component can be either discriminative models which focus on recommending existing bundles, or generative models which create new bundles.
\end{itemize}

\begin{figure}[htbp]
    \vspace{-0.3cm}
    \centering
    \includegraphics[width=0.85\linewidth]{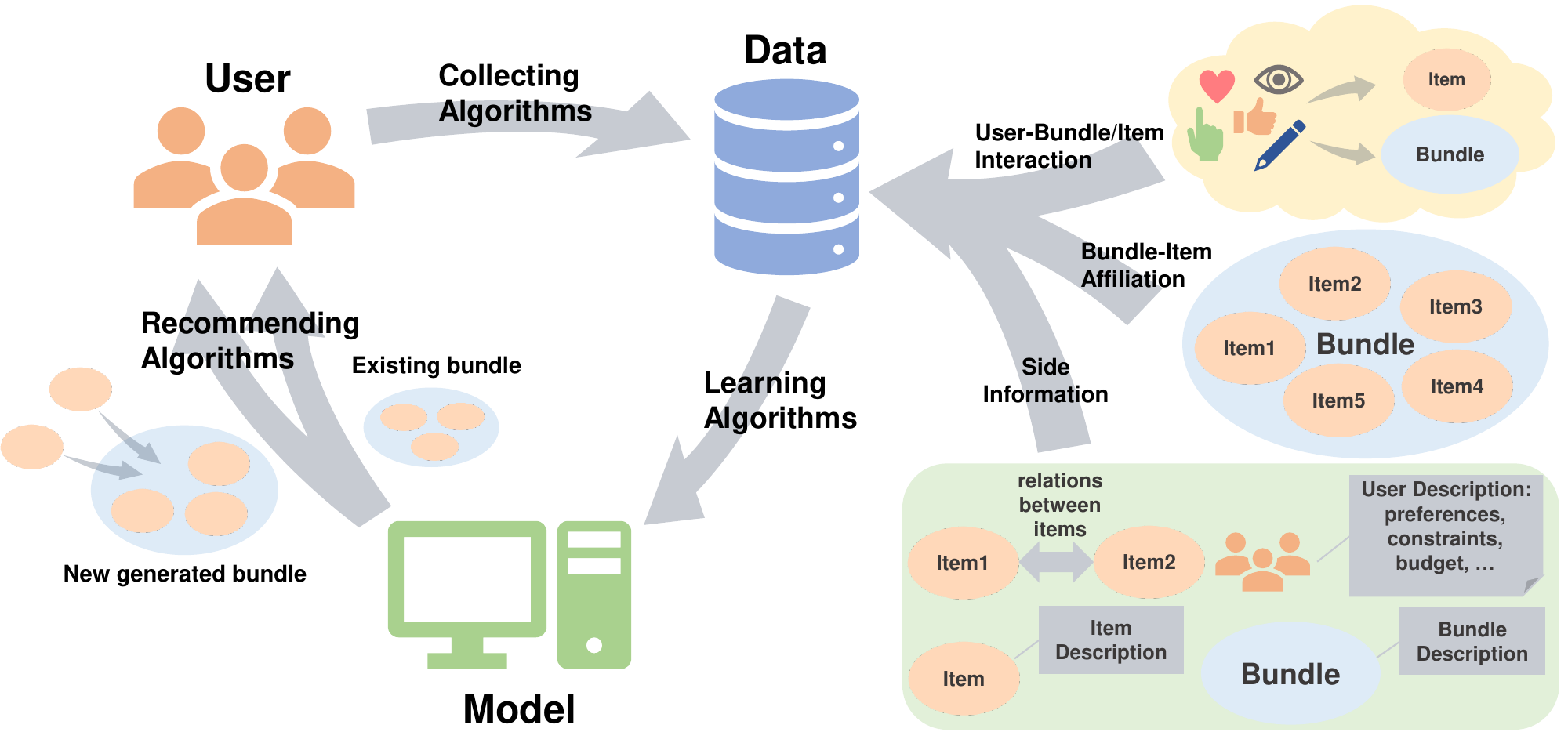}
    \vspace{-0.3cm}
    \caption{Lifecycle of Bundle Recommendation}
    \label{fig:lifecycle of BundleRS}
    \vspace{-0.5cm}
\end{figure}

As shown in Figure \ref{fig:lifecycle of BundleRS}, this entire lifecycle can be interconnected by these three components, encapsulating in three distinct stages:
\begin{itemize}
    \item \textbf{User\(\rightarrow\)Data.} This stage is to collect user behaviors towards bundles and items, including implicit feedback like browsing, clicking, and explicit feedback like rating, reviewing. These actions represent user-bundle interactions and user-item interactions. Bundle composition information (e.g., the items within each bundle) and some side information (e.g., user profile, bundle and item attributes, contexts) are also gathered.
    \item \textbf{Data\(\rightarrow\)Model.} This stage is to devise appropriate models and train the models using the collected data to learn user preferences towards bundles and items, further to capture the intra- and inter-relationship among bundles.
    \item \textbf{Model\(\rightarrow\)User.} This stage is that the model returns the recommendation or generation results to users, which usually predicts the scores for unseen user-bundle pairs, or generates new bundles based on user preferences which are then recommended to the user.
\end{itemize}

Through the entire lifecycle of bundle recommendation, user and bundle RS are in a process of  mutual dynamic evolution. User interactions and feedback, as well as dynamic changes of bundle can continuously update and improve the model. Consequently, the bundle recommendation results can influence user preferences and lead to updates in bundle composition.

\subsection{Problem Formulation}
\paragraph{\textbf{Entities and Definitions}}
Let's use $\mathcal{U} = \{u_1, u_2, ..., u_M\}$, $\mathcal{B} = \{b_1, b_2, ..., b_N \}$, and $\mathcal{I} = \{i_1, i_2, ..., i_O \}$ to denote the sets of users, bundles and items. $M$, $N$ and $O$ denote the number of users, bundles and items, respectively. Each bundle $b = \{i_1, i_2, ..., i_K \}$, $b \in \mathcal{B}$, $i \in \mathcal{I}$. $K$ is the size of bundle $b$.

\paragraph{\textbf{Interactions and Affiliations}}
Let's denote user-bundle interaction matrix, user-item interaction matrix, and bundle-item affiliation matrix as $\mathbf{X}_{M \times N} = \{\mathbf{x}_{u b}\mid u \in \mathcal{U}, b \in \mathcal{B} \}$, $\mathbf{Y}_{M \times O} = \{\mathbf{y}_{u i}\mid u \in \mathcal{U}, i \in \mathcal{I} \}$, and $\mathbf{Z}_{N \times O} = \{\mathbf{z}_{b i}\mid b \in \mathcal{B}, i \in \mathcal{I} \}$. An observed interaction $\mathbf{x}_{u b}$ between user $u$ and bundle $b$, or $\mathbf{y}_{u i}$ between user $u$ and item $i$, can either be a binary value (0 or 1) or a rating. An entry $\mathbf{z}_{b i}$ is a binary value indicating whether bundle $b$ contains item $i$. We denote the collected interaction data as $D_T$, and the collected side information data as $D_S$ including user constraints, item descriptions, relations between items and so on.

\paragraph{\textbf{Task Definition}}
Based on the above definition, the task of bundle recommendation can be stated as follows: given the interaction data $D_T$ and side information data $D_S$, learn a bundle recommendation model $f$ that can predict the likelihood of a user $u$ accepting a bundle $b$. The bundle $b$ can either be an existing bundle or a newly generated bundle. This is formalized as:
\begin{equation}
\hat{y} = f(u, b, D_T, D_S).
\end{equation}

\subsection{Taxonomy}
As shown in Figure \ref{fig:taxonomy of bundle recommendation}, we introduce a taxonomy of bundle recommendation based on bundling strategy from various application scenarios, classifying it into two distinct categories:
\begin{figure}[htbp]
\vspace{-0.3cm}
    \centering
    \includegraphics[width=0.8\linewidth]{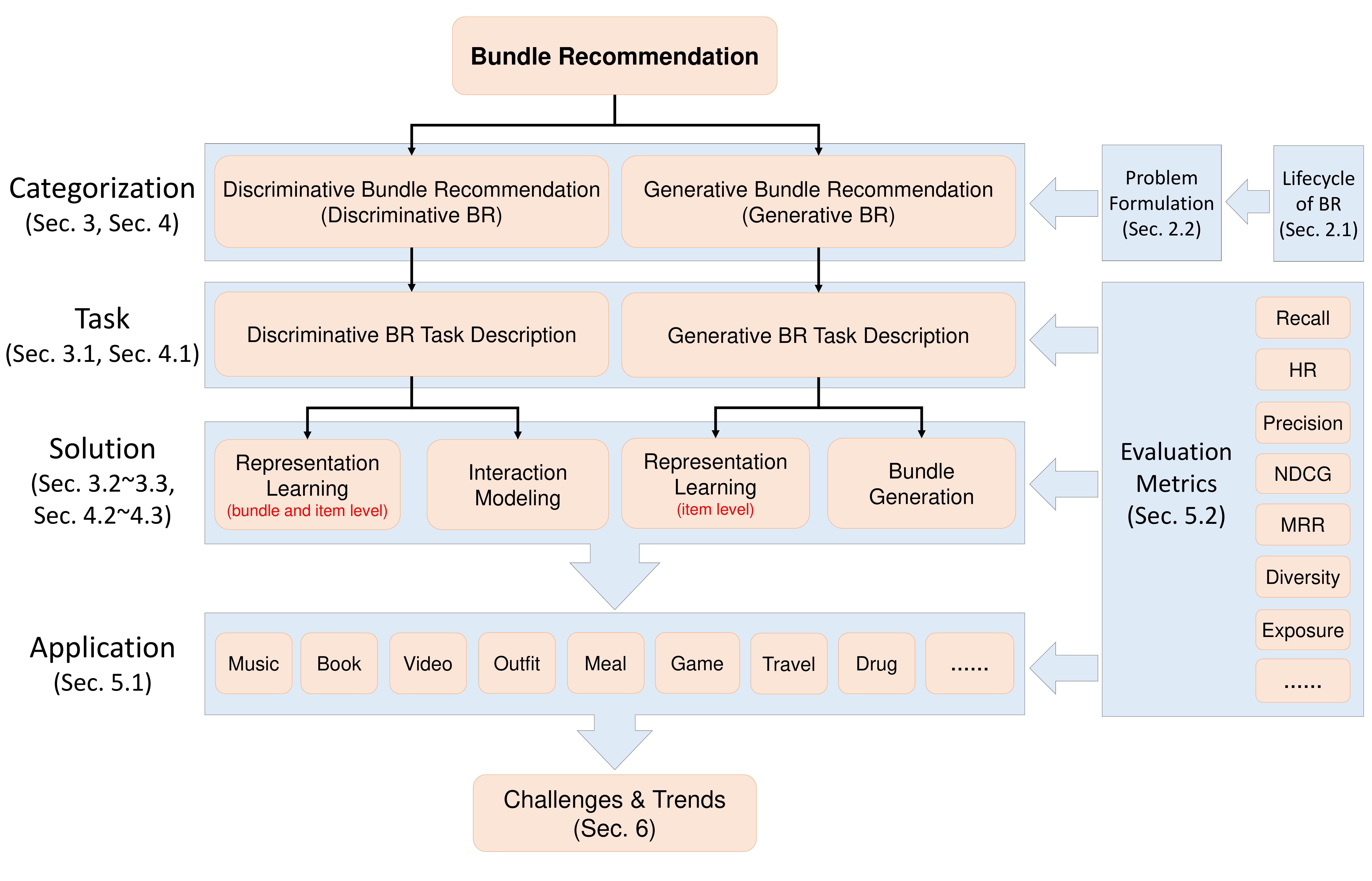}
    \vspace{-0.2cm}
    \caption{Taxonomy of Bundle Recommendation}
    \label{fig:taxonomy of bundle recommendation}
    \vspace{-0.5cm}
\end{figure}
\begin{itemize}
    \item \textbf{Discriminative Bundle Recommendation (Discriminative BR)} is aimed at scenarios where the task is to predict the likelihood that a user will favor pre-existing bundles.
    \item \textbf{Generative Bundle Recommendation (Generative BR)} is aimed at scenarios where the task is to create bundles, recommending these generated bundles to users.
\end{itemize}
The pre-existence of bundles can be co-consumed products \cite{liu2017modeling}, user-generated lists \cite{BGCN, AttList, he2020consistency} or manually pre-defined bundles by retailers \cite{BundleNet, liu2011personalized, CrossCBR}. However, pre-existing bundles may not always align with the personalized preferences of users. For instance, on music streaming platforms, users often expect recommendations for specific albums produced by artists or playlists that have been curated by other users or the platform itself. Alternatively, the systems might dynamically generate playlists based on user's musical tastes, creating a personalized collection of songs that reflect the user's mood or specific event.

Figure \ref{fig:taxonomy of bundle recommendation} illustrates the taxonomy of bundle recommendation, discriminative and generative, respectively. It outlines the respective tasks for different categories and proposes the corresponding solutions. Datasets and evaluation metrics are the application resources in bundle recommendation. The details are presented in the following sections. It is worth noting that this survey primarily focuses on general bundle recommendation tasks. Domain-specific scenarios (e.g., drug package recommendation, where requires detailed pharmacological knowledge) are only briefly mentioned but not discussed in depth, as they rely on strong domain expertise beyond the general scope of bundle recommendation systems.

\section{Discriminative Bundle Recommendation}
\label{section3:discriminative}
\begin{figure}[htbp]
    \vspace{-0.5cm}
    \centering
    \includegraphics[width=0.9\linewidth]{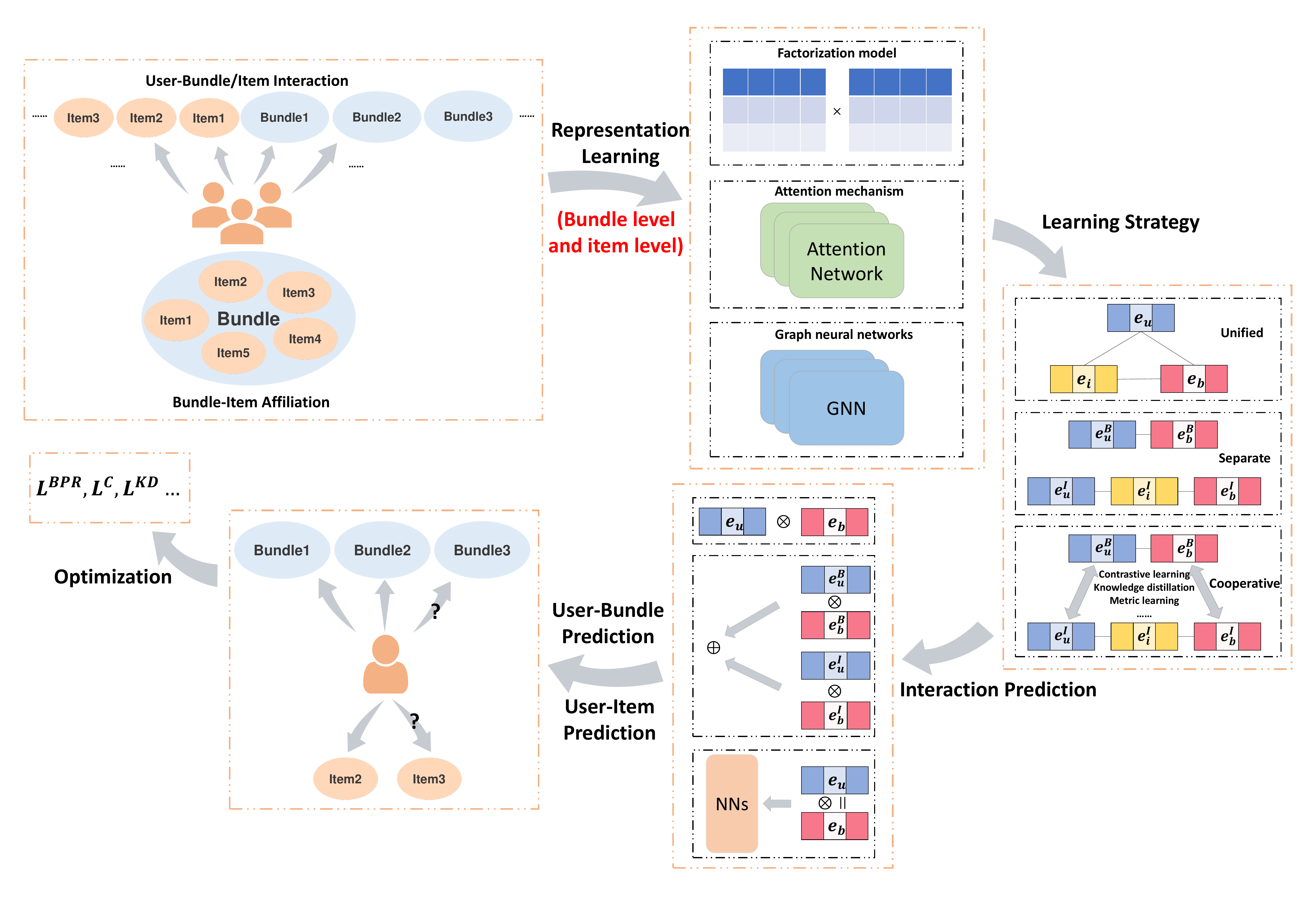}
    \vspace{-0.5cm}
    \caption{Framework of Discriminative Bundle Recommendation}
    \label{fig:framework of discriminative bundle RS}
    \vspace{-0.5cm}
\end{figure}

Discriminative bundle recommendation (Discriminative BR) aims to recommend predefined bundles such as fashion outfit, curated album, etc. It focuses on the task of identifying the most relevant and appealing bundles for a given user by leveraging discriminative models. These models are usually designed to predict the score of unseen user-bundle or user-item pair based on user's behaviors and historical interactions. The framework of Discriminative BR is illustrated in Figure \ref{fig:framework of discriminative bundle RS}. Discriminative BR research utilizes different learning strategies (unified, separate, cooperative) to learn high-quality representations of users, items and bundles from bundle and item levels. After obtaining the representations, common methods such as inner product, neural networks, or other techniques are used to predict the score of the unseen user-bundle or user-item pairs. The subsequent section will first discuss the Discriminative BR task and then explore two main parts of Discriminative BR: representation learning and interaction modeling. And finally we provide a summary of existing Discriminative BR research.

\subsection{Discriminative BR Task}
The goal of discriminative bundle recommendation task \cite{BGCN, CrossCBR, HBGCN} is to predict the likelihood or score that a user will interact with a given bundle. This can be represented by a function $f_{\text{disc}}$ which takes as input a user $u$, a bundle $b$, user's interactions with both bundles $\mathbf{X}_{M \times N}$ and individual items $\mathbf{Y}_{M \times O}$, as well as the composition of the bundles $\mathbf{Z}_{N \times O}$, and possibly other side information $c$, and outputs a predicted score $\hat{y}_{ub}$. The function can be mathematically expressed as:
\begin{equation}
\hat{y}_{ub} = f_{\text{disc}}(u, b, \mathbf{X}_{M \times N}, \mathbf{Y}_{M \times O},\mathbf{Z}_{N \times O}, c ; \theta), 
\end{equation}
where $\theta$ represents the parameters of the discriminative model. The model parameters $\theta$ are learned by minimizing the loss between the predicted and actual interactions:
\begin{equation}
\theta^* = \arg \min_{\theta} \sum_{(u,b) \in \mathcal{D}_{\text{train}}}\mathcal{L}(y_{ub}, f_{\text{disc}}(u, b, \mathbf{X}_{M \times N}, \mathbf{Y}_{M \times O}, \mathbf{Z}_{N \times O}, c ; \theta)), 
\end{equation}
where $\mathcal{L}$ is a loss function (e.g., BPR loss, cross-entropy loss), $y_{ub}$ is the actual interaction score or label, and $\mathcal{D}_{\text{train}}$ is the training dataset.

\subsection{Representation Learning}
This part is aim to learn high-quality representations of users, items, and bundles from bundle and item levels based on different learning strategies. As illustrated in Table \ref{discriminative:representation learning}, we categorize different models according to representation learning approaches and strategies. The representation approach includes factorization models \cite{EFM, LIRE, DAM, BundleBPR}, attention-based models \cite{AttList, IPRec, GANCF, DGMAE, GRAM-SMOT}, and GNNs-based models \cite{BGCN, BundleNet, CrossCBR, CMRec, MultiCBR}. We will introduce the different approaches in Section \ref{learning approach} specifically. As for learning strategy, we divide them into the following three types, namely unified representation learning, separate representation learning, and cooperative representation learning. This part will be discussed in Section \ref{learning strategy}.

\begin{table}[htbp]
\vspace{-0.2cm}
\captionsetup{justification=centering}
\caption{Representation Learning Approach and Strategy}
\vspace{-0.3cm}
\label{discriminative:representation learning}
\centering
\resizebox{\textwidth}{!}{
\begin{tabular}{|cc|c|c|c|}
\hline
\multicolumn{2}{|c|}{\diagbox{\textbf{Learning strategy}}{\textbf{Learning approach}}}                            & \textbf{Factorization model} & \textbf{Attention mechanism}                                                              & \textbf{GNNs}                                                                                              \\ \hline
\multicolumn{2}{|c|}{\textbf{Unified}}                                                    & \begin{tabular}[c]{@{}c@{}}EFM\cite{EFM}, FPITF\cite{FPITF}\\LIRE\cite{LIRE}, FPMC\cite{FPMC}\end{tabular}                   & PGAT\cite{PGAT}, AttList\cite{AttList}, IPRec\cite{IPRec}, HGLR\cite{HGLR}                                                                & \begin{tabular}[c]{@{}c@{}}BundleNet\cite{BundleNet}, PGAT\cite{PGAT}, CLBR\cite{CLBR}\\  HED\cite{HED}, UHBR\cite{UHBR}, HGLR\cite{HGLR} \\ HFGN\cite{HFGN}, AGF\cite{AGF}\end{tabular}             \\ \hline
\multicolumn{2}{|c|}{\textbf{Separate}}                                                 & -                            & \begin{tabular}[c]{@{}c@{}}BRUCE\cite{BRUCE}, DT-CDBR\cite{DT-CDBR}, MULTIPLE\cite{MULTIPLE},\\ CAPLE\cite{CAPLE}, IHBR\cite{IHBR}, BundleGT\cite{BundleGT}\end{tabular} & \begin{tabular}[c]{@{}c@{}}BGCN\cite{BGCN}, BGGN\cite{BGGN}, BundleGT\cite{BundleGT}\\ HBGCN\cite{HBGCN}, IHBR\cite{IHBR}, RGNN\cite{wang2021relational} \\ DT-CDBR\cite{DT-CDBR}, CMRec\cite{CMRec}, TGIF\cite{kim2024less}   \\ BasConv\cite{BasConv}, SuGeR\cite{SuGeR} \end{tabular} \\ \hline
\multicolumn{1}{|c|}{\multirow{6}{*}{\textbf{Cooperative}}} & -                                & DAM\cite{DAM}, BundleBPR\cite{BundleBPR}              & GANCF\cite{GANCF}, DAM\cite{DAM}                                                                                & IMBR\cite{IMBR}                                                                                                       \\ \cline{2-5} 
\multicolumn{1}{|c|}{}                                      & \textbf{Knowledge Distillation} & -                            & DGMAE\cite{DGMAE}                                                                                     & DGMAE\cite{DGMAE}                                                                                                      \\ \cline{2-5} 
\multicolumn{1}{|c|}{}                                      & \textbf{Contrastive Learning}   & -                            & -                                                                                         & \begin{tabular}[c]{@{}c@{}}GPCL\cite{GPCL}, MIDGN\cite{MIDGN}, HIDGN\cite{zou2024towards}\\ CateRec\cite{li2024boosting}, CrossCBR\cite{CrossCBR}, C3BR\cite{C3BR} \\ CL2BRec\cite{CL2BRec}, CoHEAT\cite{CoHEAT}, EBRec\cite{EBRec}\\ MultiCBR\cite{MultiCBR}, DSCBR\cite{wu2024dual}\end{tabular}               \\ \cline{2-5} 
\multicolumn{1}{|c|}{}                                      & \textbf{Metric Learning}        & -                            & GRAM-SMOT\cite{GRAM-SMOT}                                                                                 & -                                                                                                          \\ \cline{2-5} 
\multicolumn{1}{|c|}{}                                      & \textbf{Mutual Learning}        & -                            & -                                                                                         & HyperMBR\cite{HyperMBR}                                                                                                   \\ \cline{2-5} 
\multicolumn{1}{|c|}{}                                      & \textbf{Curriculum Learning}    & -                            & -                                                                                         & CoHEAT\cite{CoHEAT}                                                                                                     \\ \hline
\end{tabular}
}
\vspace{-0.5cm}
\end{table}

\subsubsection{Backbone Models of Representation Learning}
\label{learning approach}
\paragraph{\textbf{Factorization models.} }
Factorization models, a subset of machine learning models, are widely employed in various domains such as recommendation systems, natural language processing, and computer vision. These models aim to decompose high-dimensional data into lower-dimensional latent factors, capturing the underlying structure and relationships within the data.
			
In the context of bundle recommendation, factorization models can be particularly effective in learning representations from multiple perspectives: user-bundle, user-item, and bundle-item. This involves decomposing the user-bundle ($\mathbf{X}_{M \times N}$), user-item ($\mathbf{Y}_{M \times O}$), and bundle-item ($\mathbf{Z}_{N \times O}$) interaction matrices into low-dimensional latent vectors: user latent vector ($\mathbf{p}_u$), item latent vector ($\mathbf{r}_i$), and bundle latent vector ($\mathbf{q}_b$). The decomposition can be represented as follows:
\begin{equation}
    \mathbf{X}_{M \times N} = \mathbf{p}_u^T \mathbf{q}_b, \quad \mathbf{Y}_{M \times O} = \mathbf{p}_u^T \mathbf{r}_i, \quad \mathbf{Z}_{N \times O} = \mathbf{q}_b^T \mathbf{r}_i. 
\end{equation}
Previous research \cite{FPMC, min2020food} did not consider bundles as collections of items; instead, it treated bundles as atomic units, similar to individual items. Then MF-based collaborative filtering is employed to learn bundle-level user preferences from user-bundle interaction matrix. Later, the distinction between bundles and items has received increasing attention. With the development of pairwise ranking approach, some works have attempted to rank the bundle by optimizing a pairwise ranking loss. LIRE \cite{LIRE}, EFM \cite{EFM} and BundleBPR \cite{BundleBPR} simultaneously utilize the users' interactions with both items and bundles under the BPR framework \cite{rendle2009bpr}. The bundle vectors can be aggregated from the latent vectors of the items they contain. By optimizing these different latent factor vectors using the Bayesian Personalized Ranking (BPR) loss, factorization models can learn comprehensive and effective representations of both users and bundles. Furthermore, BundleBPR model \cite{BundleBPR} leverages the parameters learned through the item BPR model to estimate the preference of a user toward a bundle. Recognizing the rich signal about item co-occurrence in bundles, ProductRec \cite{ProductRec} explores the latent bundling relationships between items based on FPMC. EFM \cite{EFM} believes that discovering the relationship between bundles and their contained items can mutually reinforce the performance of user-item and user-bundle recommendation. Out of this thought, they devise embedding factorization models, which jointly factorize user-item interaction matrix, user-bundle interaction matrix, and item-item-bundle co-occurrence matrix within a unified framework. Combining neural network techniques, DAM \cite{DAM} proposes a factorized attention network, which factorizes the attention weight matrix with a low-rank model followed by a softmax operation, differing from existing designs of neural attention network. In particular, considering the heterogeneous features of items in a bundle, FPITF \cite{FPITF} proposes a functional tensor factorization model for fashion recommendation, which decomposes the high order interactions between users and the fashion items into a set of pairwise interactions in some latent space.

\paragraph{\textbf{Attention mechanism.}}
However, the more items two bundles have in common, the more similar they are, since a bundle is composed of multiple items. Therefore, it is inappropriate to treat bundles as separate columns in the user-bundle matrix to run CF models like matrix factorization. To address the compositional similarity between bundles, it is essential to aggregate item embeddings for bundle representation. Since items within a bundle have varying degrees of importance, the attention mechanism is employed to dynamically weight the importance of different items during aggregation \cite{DAM, AttList, IPRec, HGLR}. For each item $i$ in the bundle $b$, an attention score $e_{bi}$ is calculated. The attention score is computed using a combination of item and user embeddings:
\begin{equation}
  e_{bi} = \mathbf{a}^T \tanh(\mathbf{W}_1 \mathbf{v}_i + \mathbf{W}_2 \mathbf{u}_u + \mathbf{b}), 
\end{equation}
where $\mathbf{u}_u$ is the embedding vector for the user $u$, and $\mathbf{a}$, $\mathbf{W}_1$, $\mathbf{W}_2$, and $\mathbf{b}$ are learnable parameters. The attention scores are converted into attention weights $\alpha_{bi}$ using a softmax function:
\begin{equation}
   \alpha_{bi} = \frac{\exp(e_{bi})}{\sum_{j \in b} \exp(e_{bj})}.
\end{equation}
The bundle representation $\mathbf{b}_b$ is obtained by aggregating the item embeddings $\mathbf{v}_i$ weighted by their corresponding attention weights $\alpha_{bi}$:
\begin{equation}
    \mathbf{b}_b = \textstyle\sum_{i \in b} \alpha_{bi} \mathbf{v}_i.
\end{equation}

One representative work is DAM \cite{DAM} which designs a factorized attention network to aggregate the item embeddings in a bundle to obtain the bundle's representation. The design of the weighting strategy can be seen as factorizing the attention weight matrix with a low-rank model. Besides the intra-connections among the items within the bundle, there are also inter-connections among bundles. IPRec \cite{IPRec} considers the inter-connection relationships for bundle modeling, and proposes a intra- and inter-bundle attention network for bundle recommendation. Similarly, AttList \cite{AttList} models latent representations of bundles and users via item-level and bundle-level attention networks. On the other hand, the items within a bundle are usually gathered according to a specific theme, which may contain some latent attributes. GANCF \cite{GANCF} proposes item-level and attribute-level attention networks to reinforce the representations of bundles. HGLR \cite{HGLR} introduces an attribute-based heterogeneous information network for representation learning and utilizes an attention network to fuse the embeddings learned from bundle and item views.

\paragraph{\textbf{GNNs.}}
Bundle recommendation involves modeling interactions among users, items, and bundles. To capture these relations, a graph is constructed where nodes represent users, items, and bundles, and edges encode user–item interactions, user–bundle interactions, or bundle–item affiliations. Each node is initialized with an embedding vector, either randomly or from pre-trained models. A graph neural network (GNN) then updates node embeddings via message passing, aggregating information from neighbors. Formally, the update for node $i$ at layer $k+1$ is:
\begin{equation}
h_i^{(k+1)} = \sigma \left( \sum_{j \in \mathcal{N}(i)} \frac{1}{c_{ij}} W^{(k)} h_j^{(k)} \right),
\end{equation}
where $h_i^{(k)}$ is the embedding of node $i$ at layer $k$, $\mathcal{N}(i)$ its neighborhood, $c_{ij}$ a normalization coefficient, $W^{(k)}$ a learnable weight matrix, and $\sigma$ an activation function.

\begin{figure}[htbp]
\vspace{-0.3cm}
  \centering
  \includegraphics[width=0.7\linewidth]{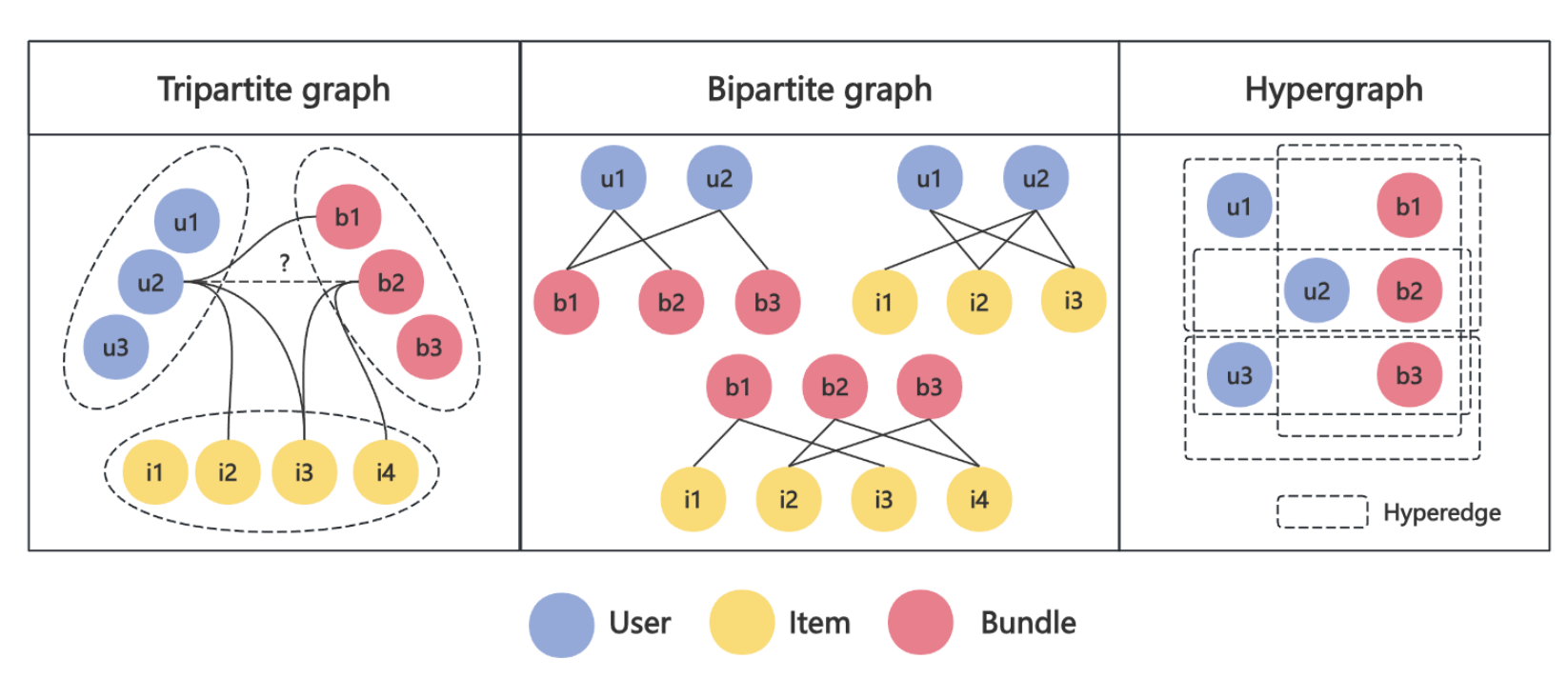}
  \vspace{-0.2cm}
  \caption{Three graph types: tripartite graph, bipartite graph, hypergraph}
  \label{graph_types}
  \vspace{-0.3cm}
\end{figure}

Motivated by influential surveys that categorize graph structures in recommender systems \cite{Gao2023survey, Anand2025survey}, we tailor this perspective to the task of bundle recommendation. We categorize existing studies into three representative graph types: tripartite graph, bipartite graph, hypergraph. Most works can be distinguished by whether users, bundles, and items are represented in a single graph: some unify all three types of nodes into a tripartite graph, while others decompose them into three bipartite graphs. While bipartite and tripartite graphs model pairwise relationships, a richer and more general perspective is offered by hypergraphs, which directly model the one-to-many relations within bundles. Figure \ref{graph_types} illustrates the three types of graph, and we summarize representative approaches for each type below.

\textbf{Tripartite graph:} A straightforward approach is to unify all three entity types—users, bundles, and items—into a tripartite graph. This structure allows for comprehensive modeling of all relationships in one unified view, enabling graph convolutional networks to aggregate information across different types of neighbors \cite{BundleNet, PGAT, BasConv, kim2024less, CLBR}. Representative works include BundleNet \cite{BundleNet}, which employs a differentiable message passing framework to effectively capture the user preferences for bundles, which can incorporate the intermediate role of items between users and bundles on the user-item-bundle tripartite graph. CL2BRec \cite{CL2BRec} constructs a user–bundle–item tripartite graph as the global structure view for data augmentation. PGAT \cite{PGAT} constructs a unified user–bundle–item tripartite graph and employs graph attention in neighborhood aggregation to assign differentiated weights to neighbors, enhancing the expressiveness of node embeddings.

\textbf{Bipartite graph:} Compared with multiple bipartite graphs, tripartite graphs are typically sparser and involve multiple heterogeneous relation types, making representation learning more challenging. In contrast, each bipartite graph contains only a single relation, leading to a denser structure and a more tractable learning process. Moreover, the separation into multiple bipartite graphs naturally aligns with the paradigm of contrastive learning, which has become an important driving force in practice. Consequently, extensive research has explored bipartite graph-based modeling for bundle recommendation \cite{BundleGT, HyperMBR, HBGCN, MultiCBR, IHBR, IMBR, DGMAE, MIDGN, CrossCBR, EBRec, CoHEAT, wu2024dual}, with most methods relying on three bipartite graphs: U-B, U-I, and B-I. Building on this framework, BGCN \cite{BGCN} unifies U–I, U–B, and B–I relations into a heterogeneous graph, where item nodes act as bridges, enabling information propagation between users and bundles allowing the learned representations to capture item-level semantics. CrossCBR \cite{CrossCBR} constructs the U–B graph as the bundle view and the U–I and B–I graphs as the item view, and then applies contrastive learning across the two views. IHBR \cite{IHBR} introduces a preference-transfer propagation mechanism over U–B and U–I graphs, while MIDGN \cite{MIDGN} disentangles user and bundle representations at the intent level by leveraging U–I and B–I graphs.\looseness=-1

\textbf{Hypergraph:} Traditional graphs are limited to pairwise relations, which cannot fully capture the one-to-many nature of bundles. Hypergraphs provide a natural solution by introducing hyperedges that can directly connect a bundle to all its constituent items \cite{DHCL-BR, HED, UHBR, yu2023basket, RaMen}. This structure allows for faithful modeling of high-order relations without decomposing them into pairwise edges. For example, HED \cite{HED} constructs a complete hypergraph to jointly model inter- and intra-connections among users, items, and bundles, while UHBR \cite{UHBR} updates embeddings on the hypergraph to leverage higher-order associations. DHCL-BR \cite{DHCL-BR} models both direct and indirect user-bundle interactions as hypergraphs and leverages dual contrastive learning to improve embedding quality. A hypergraph convolution framework is proposed for bundle representation learning, connecting all items in a bundle as hyperedges and further linking repeated items to strengthen intra-bundle correlations \cite{yu2023basket}. RaMen \cite{RaMen} is a multi-strategy, multi-modal approach for bundle construction, where hypergraph message passing captures shared latent intents among item groups. These works highlight that hypergraph-based techniques effectively model complex high-order interactions.

\subsubsection{Representation Learning Strategy}
\label{learning strategy}
\begin{figure}[htbp]
\vspace{-0.3cm}
  \centering
  \includegraphics[width=0.7\linewidth]{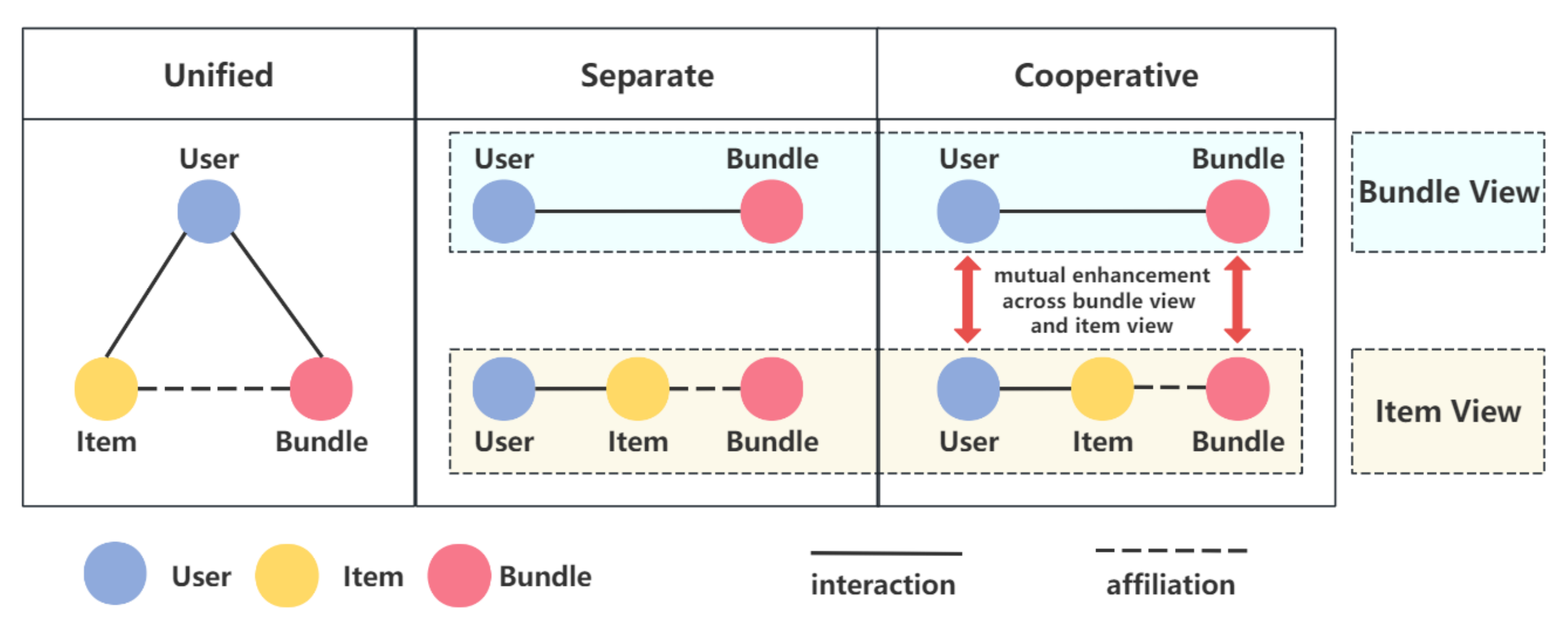}
  \vspace{-0.3cm}
  \caption{Three representation learning strategies (unified, separate, cooperative)}
  \label{Three representation learning strategies}
  \vspace{-0.5cm}
\end{figure}

\ 
\newline
Since there are U-I, U-B, and B-I interactions in bundle recommendation leading to two views, it is natural to focus on learning strategy between these two views when learning representations. Therefore, we classify the representation learning from the perspective of learning strategy between bundle view and item view. For ease of explanation, we illustrate the three representation learning strategies in Figure \ref{Three representation learning strategies}. One research line blindly merges the two views into a unified tripartite graph and employs GNNs or attention mechanism to aggregate the neighboring information into representations of users and bundles \cite{BundleNet, HED, AttList, IPRec}. We term this approach as \textbf{unified learning} strategy. Conversely, the second line performs representation learning upon the two views individually, and then fuses these two view-specific predictions \cite{BGCN, HBGCN, IHBR, CMRec}. This approach is defined as \textbf{separate learning} strategy. Finally, the third line properly models the cooperative association and encourages the mutual enhancement across the bundle and item views through techniques such as cross-view contrastive learning, metric learning, or knowledge distillation, among others \cite{CrossCBR, MultiCBR, DGMAE, HyperMBR}. We define this as \textbf{cooperative learning} strategy.
\paragraph{\textbf{Unified Learning.}}
The unified representation learning strategy aims to learn representations of users and bundles by integrating user-bundle, user-item, bundle-item interactions within a unified framework. This approach typically employs techniques such as GNNs or attention mechanism to propagate information and update embeddings across a unified structure. This structure can take forms like a user-bundle-item tripartite graph, hierarchical structure, or other complex networks.

In this unified view, user nodes $(U)$, item nodes $(I)$, and bundle nodes $(B)$ are interconnected through various interaction edges, such as user-bundle interactions, user-item interactions, and bundle-item interactions. By merging these interactions into a unified representation space, the strategy aggregates information from neighboring nodes to generate robust representations for users and bundles. This process can be expressed abstractly as:
\begin{equation}
 \mathbf{e}_u, \mathbf{e}_b \leftarrow U \text{-} B \text{-} I,    
\end{equation}
where $\mathbf{e}_u$ and $\mathbf{e}_b$ represent the user and bundle embeddings obtained through the aggregation of neighboring information within the unified structure. And the final prediction is just to fuse user embeddings and bundle embeddings:
\begin{equation}
     \hat{y} = f(\mathbf{e}_u, \mathbf{e}_b).
\end{equation}
Here, \( f \) is a prediction function.

There are some existing research studies that utilized the unified strategy, which integrated users, items, and bundles into a unified user-bundle-item graph \cite{BundleNet, PGAT, CLBR} or hypergraph \cite{HED, UHBR}, and propagated neighboring information to get the final representations of users and bundles. However, it is obvious that user-bundle interaction density is much sparser than user-item interaction density. To confront data sparsity, counterfactual data was generated for augmentation, and the unified graph-based BR model was guided to utilize this data effectively as proposed in \cite{CLBR}. Furthermore, user preference naturally propagates hierarchically from items to bundles and finally to users \cite{AttList, HFGN}. AttList \cite{AttList} applied attention networks to aggregate the item representations into a bundle representation space. Based on the aggregated bundle representations, they further aggregated them to model the latent representation of the user. HFGN \cite{HFGN} propagated information from lower levels to higher levels, that is, gathers information from item nodes to update bundle representations. This hierarchical approach complements the unified strategy by ensuring that detailed item-level information is effectively utilized to enhance user and bundle representations.

\paragraph{\textbf{Separate Learning.}}
The second representation learning strategy is the separate learning strategy. The main idea of this strategy is to first perform representation learning and preference prediction upon two views individually, and fuse two view-specific predictions to obtain the final prediction. It means that the representations of users and bundles are learned separately at bundle level and item level, while cooperative signals are only combined at the prediction level. Formally, the strategy can be described as follows:
\begin{equation}
    \mathbf{e}_u^B, \mathbf{e}_b^B \leftarrow U \text{-} B, \mathbf{e}_u^I, \; \mathbf{e}_b^I \leftarrow U \text{-} I, B \text{-} I
\end{equation}
where $\mathbf{e}_u^B$ and $\mathbf{e}_b^B$ are user and bundle embeddings learned at the bundle level, and $\mathbf{e}_u^I$ and $\mathbf{e}_b^I$ are user and bundle embeddings learned at the item level. Predictions at each level are made independently:
\begin{equation}
    \hat{y}_B = f_B(\mathbf{e}_u^B, \mathbf{e}_b^B),\; \hat{y}_I = f_I(\mathbf{e}_u^I, \mathbf{e}_b^I).
\end{equation}
Finally, the predictions from two levels are combined to obtain the final prediction:
\begin{equation}
     \hat{y} = g(\hat{y}_B, \hat{y}_I).
\end{equation}
Here, \( f_B \) and \( f_I\) are prediction functions at the bundle level and item level, respectively. And \( g \) is a function that combines these predictions to generate the final prediction.

There are several research studies \cite{BGCN, BGGN, wang2021relational, DT-CDBR, HBGCN, CMRec} etc. which propagate information at bundle level and item level, and combine bundle and item levels for final prediction. Considering that users have different preferences for different bundle categories, CMRec \cite{CMRec} adopts category-wise propagation at two levels, and combines category-wise user and bundle representations from both levels to estimate the likelihood of user-bundle interactions. Additionally, since the bundle-level and item-level preferences of the same user in the same time period are intertwined with each other, MULTIPLE \cite{MULTIPLE} leverages the bundle-item containment relationship and obtain dual-level user preferences that incorporate information at both two levels. Specially, based on learning user preferences from bundle level and item level, BundleGT \cite{BundleGT} further combines the information of bundling strategy into the user and bundle representations.

\paragraph{\textbf{Cooperative Learning.}}
The third representation learning strategy is the cooperative representation learning strategy. This approach is similar to the separate learning strategy in that it learns user and bundle representations from bundle view and item view. However, the key difference is that, techniques such as contrastive learning, knowledge distillation, metric learning are employed cross two views to achieve mutual enhancement. Formally, the strategy can be described as follows:
\begin{equation}
    \begin{array}{c}
\mathbf{e}_u^B, \mathbf{e}_b^B \leftarrow U \text{-} B \\
\tikz[baseline] \draw[->] (0,0) -- (0,0.5) node[midway, right] {}; \tikz[baseline] \draw[<-] (0,0) -- (0,0.5) node[midway, right] {}; \\
\mathbf{e}_u^I, \mathbf{e}_b^I \leftarrow U \text{-} I, B \text{-} I
\end{array},
\end{equation}
where $\mathbf{e}_u^B$ and $\mathbf{e}_b^B$ are user embeddings and bundle embeddings learned at bundle level, $\mathbf{e}_u^I$ and $\mathbf{e}_b^I$ are user embeddings and bundle embeddings learned at item level. Regarding prediction, it is similar to that of the separate strategy. To avoid redundancy, this part will not be elaborated upon here.

In cooperative strategy, mutual enhancement is achieved through various techniques. One such technique is contrastive learning, which aligns and contrasts representations across the bundle view and item view to enhance representation quality. Cross-view contrastive learning has been leveraged in several research work \cite{CrossCBR, MIDGN, CL2BRec, EBRec, C3BR, CoHEAT}. CrossCBR constructs cross-view contrastive losses to enhance the representation ability of the same user (or bundle) from different views, while decrease that of different users (or bundle). Similarly, MIDGN builds contrastive losses between global and local views. CoHEAT exploits contrastive learning to align interaction-view and affiliation-view, effectively learning latent representations. The aforementioned works is to use cross-view contrastive loss to achieve cross-view cooperation. In contrast to the works mentioned above, MultiCBR \cite{MultiCBR} first fuses the multi-view representations into a unified one, and then adopts a simple self-supervised contrastive loss on the unified representations. Whether considering cross-views or multi-views, contrastive learning guarantees mutual enhancement of the various views. However, there exists a critical \textquotesingle sampling bias\textquotesingle\ issue that similar nodes might be pushed far apart, leading to performance degradation. GPCL \cite{GPCL} addresses this by devising a prototypical contrastive objective to capture the correlations between a user/bundle and its prototype. Self-supervised contrastive learning has well advanced the development of bundle recommendation. However, it may misclassify some positive samples as negative samples, resulting in suboptimal models. Inspired by this idea, DSCBR \cite{wu2024dual} integrates supervised and self-supervised contrastive learning to exploit the full potential of contrastive learning for bundle recommendation.

Additionally, knowledge distillation (KD) can be employed to transfer knowledge between two views, further improving representation learning ability. KD was proposed for model compression by transferring knowledge from a large teacher model to a smaller, lightweight student model. Some studies have applied KD to RS \cite{kang2020rrd, kweon2021bidirectional, lee2019collaborative}. In bundle recommendation, the additional U-I interactions and B-I affiliation information allow to decompose the bundle into finer-grained items and infer the reason for user's interaction with bundles at the item level. For instance, DGMAE \cite{DGMAE} uses knowledge distillation to transfer the knowledge from bundle view to item view for more fine-grained learning of user's bundle preferences.

Building upon the concept of knowledge transfer, another effective technique is mutual learning, which optimizes the distance metrics to ensure that similar users or bundles are closer in the representation space. Unlike KD, which usually involves a unidirectional transfer from teacher to student, mutual learning involves the simultaneous exchange of knowledge among multiple models during training process. This bidirectional flow of information can lead to mutual enhancement of related models. For example, HyperMBR \cite{HyperMBR} utilizes mutual learning loss to enable the representations of the two views can mutually enhance each other.

In particular, there exists a research line, which leverages shared layers \cite{DAM, GANCF} or shared parameters \cite{BundleBPR} to achieve cooperation. DAM jointly models user-bundle and user-item interactions through sharing CNN layers. And BundleBPR model leverages the parameters learned through the item BPR model to estimate the preference of a user toward a bundle.

\subsection{Interaction Modeling}
After representation learning, we obtain the learned embeddings of  user $ u $, bundle $b$ and item $i$, denoting $\mathbf{p}_u$, $\mathbf{q}_b$ and $\mathbf{r}_i$, respectively. This part focuses on interaction function modeling, which aims to estimate the user's preference for the bundle based on these representations. As summarized in Table \ref{discriminative:interaction modeling}, we categorize interaction modeling into three types: inner product, distance modeling, and neural network. For each type, we describe how to model users' predicted preference, denoted as $\hat{y}_{ub}$, based on the learned embeddings. 
\begin{table}[htbp]
\vspace{-0.3cm}
\captionsetup{justification=centering}
\caption{Interaction Modeling Techniques}
\vspace{-0.2cm}
\label{discriminative:interaction modeling}
\centering
\scalebox{0.69}{
\begin{tabular}{ccc|cccl}
\toprule
\textbf{Category}                       & \textbf{Modeling Summarization}                                                                            & \textbf{References}                                                                                                                                                                                                                                                              & \textbf{Category}                         & \textbf{Modeling Summarization}                                                                                                                                               & \textbf{References}                                                                                                                                                                                               &                      \\ \cline{1-6}
\multirow{3}{*}{\textbf{Inner Product}} & $ \hat{y}_{ub} = \mathbf{p}_u^T \cdot aggregator(\mathbf{r}_i )$                                           & \cite{BundleBPR, HFGN}                                                                                                                                                                                                                                          & \textbf{Distance Modeling}                & \begin{tabular}[c]{@{}c@{}}Hyperbolic Distance \\ $ \hat{d}_{ub} = d_\mathbb{H}(\mathbf{p}^I_u, \mathbf{q}^I_b) + d_\mathbb{H}(\mathbf{p}^B_u, \mathbf{q}^B_b) $\end{tabular} & \cite{HBGCN, HyperMBR}                                                                                                                                                                           &                      \\ \cline{2-6}
                                        & $ \hat{y}_{ub} = \mathbf{p}_u^T \mathbf{q}_b $                                                             & \begin{tabular}[c]{@{}c@{}}\cite{AGF, HED, UHBR, CAPLE},\\ \cite{BundleGT, STR-VGAE} et al.\end{tabular}                                                                                                                                       & \multirow{2}{*}{\textbf{Neural Networks}} & $ \hat{y}_{ub} = MLP(\mathbf{p}_u \| \mathbf{q}_b) $                                                                                                                          & \begin{tabular}[c]{@{}c@{}}\cite{BRUCE, IMBR, IHBR, IPRec},\\ \cite{DAM, BundleNet, AttList, kim2024less},\\ \cite{GRAM-SMOT, BundlesSEAL} et al.\end{tabular} &                      \\ \cline{2-3} \cline{5-6}
                                        & $ \hat{y}_{ub} = {\mathbf{p}^{L1}_u}^\top \mathbf{q}^{L1}_b + {\mathbf{p}^{L2}_u}^\top \mathbf{q}^{L2}_b $ & \begin{tabular}[c]{@{}c@{}}\cite{CoHEAT, BasConv, BGCN, MIDGN}, \\ \cite{C3BR, CrossCBR, DT-CDBR, DGMAE}, \\ \cite{CL2BRec, GPCL, zou2024towards},\\ \cite{CMRec, EBRec, li2024boosting} et al.\end{tabular} &                                           & $ \hat{y}_{ub} = CNN(\mathbf{p}_u \otimes \mathbf{q}_b) $                                                                                                                     & \cite{GANCF}                                                                                                                                                                                     & \multicolumn{1}{c}{} \\ \bottomrule
\end{tabular}
}
\end{table}

\subsubsection{Inner Product}
To estimate the unseen user-bundle pair score, a straightforward method involves using inner product between the learned user embeddings and bundle embeddings. The equation for this is as follows:
\begin{equation}
    \hat{y}_{ub}=\mathbf{p}_u^T \mathbf{q}_b.
\end{equation}
This approach is commonly utilized in various research works \cite{AGF, HED, UHBR, CAPLE}. The bundle embeddings can be learned directly from user-bundle interactions, also can be aggregated from item embeddings using mean pooling, sum pooling or other operations \cite{BundleBPR, HFGN}. This prediction method can be formulated as follows:
\begin{equation}
    \hat{y}_{ub} = \mathbf{p}_u^T \cdot aggregator(\mathbf{r}_i).
\end{equation}
However, due to the existence of bundle view and item view, some studies \cite{BGCN, CL2BRec, CrossCBR, MIDGN, CMRec} learn user and bundle representations from multiple views or multiple levels. Consequently, it is essential to combine different levels of predictions for the final prediction. The most commonly used method in bundle recommendation for final prediction involves first using inner product between user and bundle representations, and then combining bundle and item levels. The abstract equation is:
\begin{equation}
    \hat{y}_{ub} = {\mathbf{p}^{L1}_u}^\top \mathbf{q}^{L1}_b + {\mathbf{p}^{L2}_u}^\top \mathbf{q}^{L2}_b.
\end{equation}
Here, $\mathbf{p}^{L1}_u$ and $\mathbf{p}_u^{L2}$ denote item-level and bundle-level user embeddings, or denote local-level and global-level user embeddings, respectively, while $\mathbf{q}_b^{L1}$ and $\mathbf{q}_b^{L2}$ denote item-level and bundle-level bundle embeddings, or denote local-level and global-level bundle embeddings, respectively.

\subsubsection{Distance Modeling}
Despite the great success and simplicity of using inner product for recommendation systems, prior research suggests that simply conducting inner product would have limitations. The distance in Euclidean space can be defined and calculated using inner product. The inner product provides the basic algebraic tool for Euclidean space, allowing us to quantify the relationships between user and bundle latent vector representations and measure their distances. However, according to Boolean's theorem \cite{linial1995geometry}, Euclidean space is difficult to obtain relatively low distortion for tree-like data structures. Therefore, bundle recommendation may be more suitable for modeling in non-Euclidean space like hyperbolic space. In order to fully leverage the advantage of hyperbolic space, recent methods \cite{HBGCN, HyperMBR} propose transforming the Euclidean input features into hyperbolic embeddings, and using hyperbolic distance between user and bundle embeddings to estimate the probability of user-bundle interactions. The probability is defined as follows:
\begin{equation}
    \hat{d}_{ub} = d_\mathbb{H}(\mathbf{p}^I_u, \mathbf{q}^I_b) + d_\mathbb{H}(\mathbf{p}^B_u, \mathbf{q}^B_b),
\end{equation}
where $d_\mathbb{H}(x,y)$ represents the hyperbolic distance between $x$ and $y$, which will be used to evaluate the probability of interactions between users and bundles. Here, it is composed of bundle level and item level hyperbolic distances.

\subsubsection{Neural Networks}
Inner product models linear interaction but might fail to capture complex relationships between users and bundles \cite{NCF}. Neural networks are widely recognized as powerful function approximators in recommender systems, capable of modeling complex nonlinear interactions \cite{WideDeep, DeepFM, NCF, DIN}. Following this trend, recent studies adopt neural architectures, including MLP and CNN, to mine nonlinear patterns of user-bundle interactions.

Neural Collaborative Filtering (NCF) is proposed to model the interaction function between each user-item pair using MLPs, achieving notable performance improvements. Expanding this approach to bundle recommendation, many studies \cite{AttList, BRUCE, IMBR, DAM} etc. employ MLPs to model the user-bundle interactions. The general form is abstracted as follows:
\begin{equation}
    \hat{y}_{ub} = MLP(\mathbf{p}_u \| \mathbf{q}_b).
\end{equation}
Researchers proposed leveraging CNN-based architecture for interaction modeling \cite{GANCF}. This method first performs an outer product on user and bundle embeddings to obtain the interaction map. Above the interaction map, a stack of CNN layers is placed, enabling the model to capture the nonlinear and higher-order correlations among users and bundles. The formula is as follows:
\begin{equation}
    \hat{y}_{ub} = CNN(\mathbf{p}_u \otimes \mathbf{q}_b).
\end{equation}
These nonlinear modeling approaches can capture higher-order correlations among representation dimensions. However, such improvements on performance come at the cost of increasing model complexity and computational time.

\subsection{Summary of Existing Discriminative BR Research}
As presented in Table \ref{summary of DBR}, we systematically summarize and categorize the existing Discriminative BR research works in terms of two dimensions: the input types of data used for representation learning and the goals of interaction prediction. The data types used for representation learning are categorized into user-bundle interaction, user-item interaction, bundle-item affiliation, and side-information such as attributes, cost, etc. Interaction prediction is tailored to specific goals, which we classify into user-bundle pair prediction, user-item pair prediction, and user interaction prediction for both bundle and item. This summary not only echoes the previous two subsections, but also provides a overview and categorical summary of existing Discriminative BR research.
\begin{table}[htbp]
\vspace{-0.3cm}
\captionsetup{justification=centering}
\caption{Summary of discriminative bundle recommendation research}
\vspace{-0.4cm}
\label{summary of DBR}
\centering
\resizebox{0.92\textwidth}{!}{
\begin{tabular}{|cccc|cc|c|}
\hline
\multicolumn{4}{|c|}{\textbf{Learning}}                                                                                                                                                                                                                                                                                                                                                                       & \multicolumn{2}{c|}{\textbf{Prediction}}                           & \multirow{2}{*}{\textbf{References}}                                                                                                                                                                                                                                \\ \cline{1-6}
\multicolumn{1}{|c|}{\textbf{\begin{tabular}[c]{@{}c@{}}User-Bundle\\ Interaction\end{tabular}}} & \multicolumn{1}{c|}{\textbf{\begin{tabular}[c]{@{}c@{}}User-Item\\ Interaction\end{tabular}}} & \multicolumn{1}{c|}{\textbf{\begin{tabular}[c]{@{}c@{}}Bundle-Item\\ Affiliation\end{tabular}}} & \textbf{Side Information} & \multicolumn{1}{c|}{\textbf{Bundle}}               & \textbf{Item} &                                                                                                                                                                                                                                                                     \\ \hline
\multicolumn{1}{|c|}{\mycheckmark}                                                               & \multicolumn{1}{c|}{\mycheckmark}                                                             & \multicolumn{1}{c|}{\mycheckmark}                                                               & -                                                                                                        & \multicolumn{1}{c|}{\multirow{8}{*}{\mycheckmark}} & -             & \begin{tabular}[c]{@{}c@{}}\cite{BundleNet, BGCN, MIDGN, BundleGT, CL2BRec, CrossCBR}\\ \cite{GPCL, CLBR, HBGCN, HyperMBR, DGMAE, SuGeR}\\ \cite{BRUCE, BGGN, EBRec, CoHEAT, PopCon, MultiCBR, IHBR}\\ \cite{AGF, PGAT, GRAM-SMOT, DT-CDBR, HED, UHBR}\end{tabular} \\ \cline{1-4} \cline{6-7} 
\multicolumn{1}{|c|}{\mycheckmark}                                                               & \multicolumn{1}{c|}{\mycheckmark}                                                             & \multicolumn{1}{c|}{\mycheckmark}                                                               & -                                                                                                        & \multicolumn{1}{c|}{}                              & \mycheckmark  & \cite{DAM, IMBR, wang2021relational}                                                                                                                                                                                                                                \\ \cline{1-4} \cline{6-7} 
\multicolumn{1}{|c|}{\mycheckmark}                                                               & \multicolumn{1}{c|}{\mycheckmark}                                                             & \multicolumn{1}{c|}{\mycheckmark}                                                               & \textbf{Category, Attributes, Genre}                                                                                             & \multicolumn{1}{c|}{}                              & -             & \cite{C3BR, HGLR, CMRec, li2023user}                                                                                                                                                                                                                                \\ \cline{1-4} \cline{6-7} 
\multicolumn{1}{|c|}{\mycheckmark}                                                               & \multicolumn{1}{c|}{-}                                                                        & \multicolumn{1}{c|}{\mycheckmark}                                                               & -                                                                                                        & \multicolumn{1}{c|}{}                              & -             & \cite{AttList, HFGN}                                                                                                                                                                                                                                  \\ \cline{1-4} \cline{6-7} 
\multicolumn{1}{|c|}{\mycheckmark}                                                               & \multicolumn{1}{c|}{\mycheckmark}                                                             & \multicolumn{1}{c|}{-}                                                                          & -                                                                                                        & \multicolumn{1}{c|}{}                              & -             & \cite{EFM, LIRE, CAPLE, MULTIPLE, BundleBPR}                                                                                                                                                                                                                        \\ \cline{1-4} \cline{6-7} 
\multicolumn{1}{|c|}{\mycheckmark}                                                               & \multicolumn{1}{c|}{\mycheckmark}                                                             & \multicolumn{1}{c|}{-}                                                                          & -                                                                                                        & \multicolumn{1}{c|}{}                              & \mycheckmark  & \cite{GANCF}                                                                                                                                                                                                                                                        \\ \cline{1-4} \cline{6-7} 
\multicolumn{1}{|c|}{\mycheckmark}                                                               & \multicolumn{1}{c|}{-}                                                                        & \multicolumn{1}{c|}{-}                                                                          & -                                                                                                        & \multicolumn{1}{c|}{}                              & -             & \cite{IPRec, STR-VGAE, NATR}                                                                                                                                                                                                                                \\ \cline{1-4} \cline{6-7} 
\multicolumn{1}{|c|}{\mycheckmark}                                                               & \multicolumn{1}{c|}{-}                                                                        & \multicolumn{1}{c|}{-}                                                                          & \textbf{Category, Cost, Social Relationship}                                                                                             & \multicolumn{1}{c|}{}                              & -             & \cite{he2016socotraveler}                                                                                                                                                                                                                                           \\ \hline
\end{tabular}
}
\vspace{-0.5cm}
\end{table}

\section{Generative Bundle Recommendation}
\label{section4:generative}
\begin{figure}[htbp]
    \centering
    \vspace{-0.3cm}
    \includegraphics[width=0.85\linewidth]{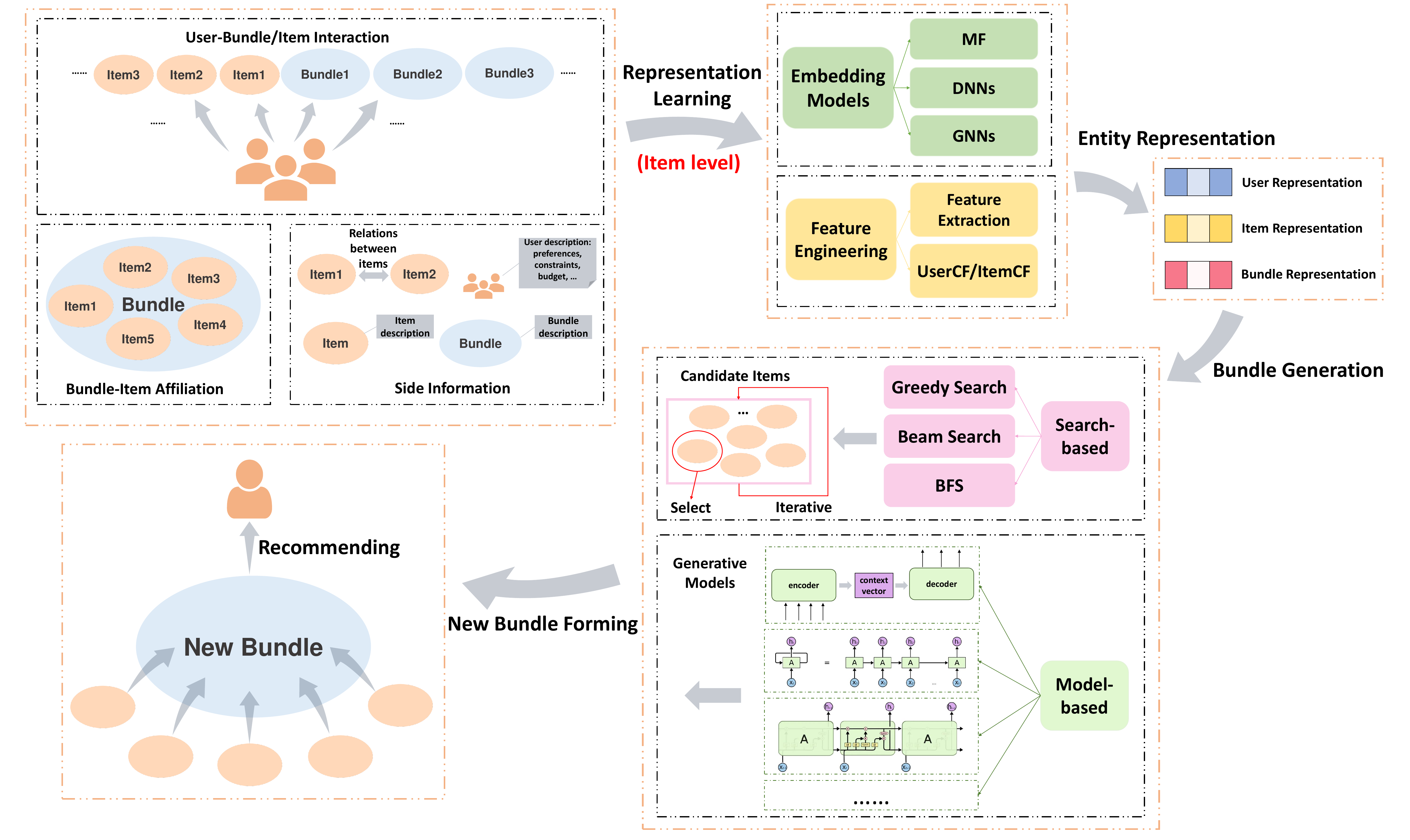}
    \vspace{-0.3cm}
    \caption{Framework of Generative Bundle Recommendation}
    \label{fig:framework of generative bundle RS}
    \vspace{-0.5cm}
\end{figure}

Unlike Discriminative BR that recommends predefined bundles, generative bundle recommendation (Generative BR) aims to create new bundles for users based on generative models. These models, often based on neural networks and probabilistic frameworks, are capable of producing novel combinations of items. The framework of Generative BR is illustrated in Figure \ref{fig:framework of generative bundle RS}. The process begins with representation learning and capturing relations, where embedding models are trained to embed user, item, or bundle (if exists), or using some side-information such as item value to describe items. Subsequently, generative models can be devised to directly create bundles such as sequence generation based models or graph generation based models, which can capture high-order relations between items and user preferences. Then, search strategies such as greedy search, beam search or others can be used to iteratively select items by maximizing utility functions. This section will be divided into four parts: generative task, representation learning from item level, bundle generation, and summary of existing Generative BR research, respectively.

\subsection{Generative BR Task}
The goal of generative bundle recommendation task \cite{BGGN, BGN, GRAM-SMOT} is to generate new bundles that a user might be interested in. This can be represented by a function $g_{\text{gen}}$ which takes as input a user $u$, and possibly a bundle $b$, user's interactions with both bundles $\mathbf{X}_{M \times N}$ and individual items $\mathbf{Y}_{M \times O}$, as well as the composition of the bundles $\mathbf{Z}_{N \times O}$, and other side information $c$, and outputs a new bundle $b'$. The function can be mathematically expressed as:
\begin{equation}
    b' = g_{\text{gen}}(u, b, \mathbf{X}_{M \times N}, \mathbf{Y}_{M \times O}, \mathbf{Z}_{N \times O}, c;  \phi),
\end{equation}
where $\phi$ represents the parameters of generative model. The function $g_{\text{gen}}$ could be a generative model such as RNN, GAN, or a more traditional method like specifying hard constraints. The learning process for a generative model typically involves maximizing the likelihood of the generated bundles given the user's interaction history. The optimization problem can be expressed as:
\begin{equation}
    \phi^* = \arg \max_{\phi} \textstyle\sum_{u \in \mathcal{U}_{\text{train}}} \log p(g_{\text{gen}}(u, b, \mathbf{X}_{M \times N}, \mathbf{Y}_{M \times O}, \mathbf{Z}_{N \times O}, c;  \phi) \mid u),
\end{equation}
where $p(\cdot \mid u)$ is the probability distribution of the generated bundles given a user $u$. This distribution reflects how likely different bundles are to be generated for user $u$ based on the model. And $\mathcal{U}_{\text{train}}$ is the training dataset.

\subsection{Representation Learning from Item Level}
\label{generative:representation learning}
When generating new bundle, we should decide which items should be included in the final bundles. So it is necessary to learn representations of entities in order to capture user preference, complex relations between entities. After surveying some related papers, we categorize the methods into two lines, as shown in Table \ref{generative:representation learning from item level}. One line is to train embedding models to represent users, items. The other line is to use side information to represent users and items, such as item values, cost, popularity, user constraints, etc.

\begin{table}[htbp]
\vspace{-0.3cm}
\captionsetup{justification=centering}
\caption{Representation Learning Methods}
\vspace{-0.3cm}
\label{generative:representation learning from item level}
\centering
\scalebox{0.68}{
\begin{tabular}{ccc|lcc}
\toprule
\textbf{Category}                          & \textbf{Summarization}      & \textbf{References}                                                                                                                                                                                                                                                                                                                                                                                             & \multicolumn{1}{c}{\textbf{Category}}         & \textbf{Summarization}                                                                                   & \textbf{References}                                                                                                                                                                                                                                       \\ \hline
\multirow{3}{*}{\textbf{Embedding Models}} & MF-based                    & \cite{yang2024nonautoregressive, zhang2024encoder}                                                                                                                                                                                                                                                                                                                                             & \multirow{3}{*}{\textbf{Feature Engineering}} & \multirow{2}{*}{\begin{tabular}[c]{@{}c@{}}Feature Extraction\\ (cost, time, price et al.)\end{tabular}} & \multirow{2}{*}{\begin{tabular}[c]{@{}c@{}}\cite{yu2015personalized, fang2014trip, chen2015package, xie2010breaking}\\ \cite{benouaret2017recommending, benouaret2016package, banerjee2020boxrec, RANDOM}\end{tabular}} \\ \cline{2-3}
                                           & \begin{tabular}[c]{@{}c@{}}DNN-based\\ (RNN, LSTM et al.)\end{tabular} & \begin{tabular}[c]{@{}c@{}}\cite{BGN, wei2022towards, han2017learning, POG, li2017mining, deng2023multiview, wang2023query2trip}\\ \cite{zheng2023interaction, CSBR, liu2023dysr, API-Prefer, deng2021build, deng2023multiaspect, CATHI} \\ \cite{ma2024leveraging, ma2024cirp, xu2024diffusion, zhang2024encoder, liu2024harnessing, DeepTrip}\end{tabular} &                                               &                                                                                                          &                                                                                                                                                                                                                                                           \\ \cline{2-3} \cline{5-6} 
                                           & \begin{tabular}[c]{@{}c@{}}GNN-based\\ (GCN, GAT et al.)\end{tabular}  & \begin{tabular}[c]{@{}c@{}}\cite{yang2024nonautoregressive, GRAM-SMOT, BGGN, su2024hierarchical}\\ \cite{zheng2023interaction, API-PROGRAM, deng2021build, li2023next}\\ \cite{ma2024leveraging, ding2023personalized, yu2023basket, zhu2023text2bundle}\end{tabular}                                                                                        &                                               & \begin{tabular}[c]{@{}c@{}}CF-based\\ (UserCF, ItemCF)\end{tabular}                                      & \begin{tabular}[c]{@{}c@{}}\cite{yu2015personalized, fang2014trip, xie2014generating}\\ \cite{beladev2016recommender, benouaret2016package, benouaret2017recommending}\end{tabular}                                     \\ \bottomrule
\end{tabular}
}
\vspace{-0.5cm}
\end{table}

\subsubsection{Embedding Models}
Similar with representation learning of Discriminative BR, both DNN-based and GNN-based models are frequently utilized in Generative BR to learn representations in many studies \cite{BGGN, GRAM-SMOT, zheng2023interaction, yang2024nonautoregressive, BGN, POG}.

\textbf{MF-based.}
BundleNAT \cite{yang2024nonautoregressive} introduces two kinds of signal to learn the needed dependency information, i.e., the preference signal and the item compatibility signal. It employs MF to extract the preference signal from user-item interactions and uses GNNs to capture item compatibility signal from the co-occurrence graph, which can strengthen the intra-relatedness within the bundle. MatTrip \cite{zhang2024encoder} utilizes Funk-SVD for feature completion to extract and complete the user's interest.

\textbf{DNN-based.}
BGN \cite{BGN} uses CNN as an encoder to extract user context information, embedding user based on user purchase history. Conna \cite{wei2022towards} adopts self-attention to learn the representations for mixture types of candidate objects. Users' interests are usually distributed across multiple aspects and a fine-grained representation learning of user interests from multiple aspects is conducted by \cite{deng2023multiaspect}. However, this approach does not fully consider the comprehensiveness and dynamism of user interests. To address this limitation, a multi-view multi-aspect neural network representation learning method is introduced by \cite{deng2023multiview}, which can more accurately model the multi-aspect interests of users and enhance the accuracy of next-basket recommendations. In works such as \cite{POG, han2017learning, li2017mining, ma2024leveraging, ma2024cirp}, items are represented using a multi-modal embedding model based on multimodal inputs, including image, text, category of the items. Recent advancements in large-scale multimodal foundation models, such as BLIP \cite{li2022blip} and CLAP \cite{wu2023large}, have paved the way for studies like CLHE \cite{ma2024leveraging} and Bundle-LLM \cite{liu2024harnessing} to harness these models for extracting the multimodal features of items. These studies have introduced self-attention mechanism to fuse multimodal information and multi-item representation. Furthermore, Bundle-LLM \cite{liu2024harnessing} employs CF method to obtain relational features from user-item interactions and bundle-item affiliations, which can enrich the representation of items. CIRP \cite{ma2024cirp} learns high-quality representations of item with multimodal encoders and pre-trained targets including image-text contrastive loss and cross-item contrastive loss. In order to capture the compatibility of items, a masked item prediction task based on a bidirectional Transformer encoder architecture is devised by \cite{POG} to capture the item interactions in a bundle. Bi-LSTM is utilized in \cite{han2017learning} to learn item compatibility within a bundle. Similarly, MatTrip \cite{zhang2024encoder} processes user interest and geographic information separately by two independent Bi-LSTM encoders. In the domain of service package generation, the process begins with parsing textual descriptions of services and mashups to distill semantic features. This is achieved through techniques such as Latent Dirichlet Allocation (LDA) and Natural Language Processing (NLP) methods, which convert text into vector representations that capture essential topics and semantics, as discussed in the works \cite{CSBR} and \cite{liu2023dysr}. Services and mashups are initially represented as static embeddings, using pre-trained word embeddings like Word2Vec or GloVe to encapsulate their inherent attributes and user requirements \cite{API-Prefer, API-PROGRAM}.

\textbf{GNN-based.}
To address the dynamic nature of services, a dynamic graph neural network is constructed to model the evolving relationships between services. The graph typically includes nodes for services, mashups, categories, and sometime words, with edges defined by various relationships such as co-occurrence, usage frequency, and semantic associations \cite{liu2023dysr}. Graph convolution operations, potentially involving GraphTransformer or Graph Convolutional Networks (GCNs), are applied to update the static embeddings by aggregating information from neighboring nodes in the graph. This allows for the dynamic adjustment of service representations in accordance with the network's topology \cite{API-Prefer, API-PROGRAM}. The graph attention mechanism is leveraged in \cite{GRAM-SMOT} to capture the high-order relationship between users, items and bundles, which can understand the constitution of a bundle better. BGGN \cite{BGGN} learns user and item embeddings through GNNs, and uses GCN to learn the complex interactions upon re-constructed bundle graph. Hypergraph convolution operations are performed on hypergraph to learn better and intent-aware representations of hyperedges \cite{yu2023basket, li2023next}. TOG \cite{ding2023personalized} first defines category combinations in outfit as templates, building two bipartite graphs (user-template and user-item), and then employ GNNs to obtain the enhanced representations of users, templates and items. In drug package scenario, DPG \cite{zheng2023interaction} models item relationships by constructing item interaction graph based on medical records and domain knowledge. It devises a message-passing neural network to capture drug interaction (synergism, antagonism, etc.) and uses DNNs to obtain the patient embeddings.

\subsubsection{Feature Engineering}
The other line of representation learning is using feature engineering techniques to represent users, items and bundles (if exists) based on some information such as item value, cost, popularity, user constraint, etc.

\textbf{Feature Extraction.}
It is common that using attributes to describe entities. For instance, each Point of Interest (POI) is depicted by a vector that includes attributes like geographic coordinates, price, and visit duration, alongside qualitative features such as category \cite{yu2015personalized, fang2014trip, benouaret2016package, benouaret2017recommending}. A part of the taxonomy is used to represent the categories of POIs in \cite{benouaret2017recommending, benouaret2016package}. FCPG \cite{chen2015package} extracts features of bundles from travelogues with TT (Term Transition) algorithm \cite{ma2013makm}, and thus forms a composite feature space composed of the bundle features and attribute features (location, time, etc.). Furthermore, each item can be represented by a value and a cost associated with it, and the user specifies a maximum total cost (budget) for any recommended set of items \cite{xie2010breaking}. Similarly in fashion outfit generation, BOXREC \cite{banerjee2020boxrec} integrates item types, price-ranges, budget, and user feedback to represent user preference, and then creates all possible three-piece clothing combinations.

\textbf{CF-based.}
Collaborative filtering (CF) is a pivotal technique in RS that analyzes similarities between users or items and leverages the preferences of similar users or characteristics of similar items to recommend items that a user may like. Across the surveyed literature, a subset of them leverages CF to model user-item interactions, where data is represented through user profiles and item features that encapsulate historical preferences and interactions.

In the context of personalized travel package recommendations, Yu et al. leverage CF to dynamically extract user preferences from location-based social network data, constructing user profiles that reflect individual preferences for various POIs \cite{yu2015personalized}. Similarly, Fang et al. propose a framework that integrates both POIs and travel packages, employing a score inference model to personalize trip recommendations, which takes into account user-based preferences and temporal properties, inferred through CF techniques \cite{fang2014trip}. Benouaret et al. further extend the application of CF by designing a package recommendation framework for trip planning. Their system emphasizes the importance of diversity and relevance in package recommendations, using a composite recommendation system inspired by CF to group diverse POIs into packages that align with user preferences and budget constraints \cite{benouaret2016package}. Building on this, Benouaret et al. aim to recommend top-k packages to users, with each package constituted by a set of POIs under specified constraints. They introduce a scoring function that considers user preferences, diversity, and popularity, all of which are integral components of CF-based representation learning \cite{benouaret2017recommending}.

Feature engineering as a data mining-driven approach to representation learning in generative bundle recommendation allows for a nuanced understanding of both the users and the items.

\subsection{Bundle Generation}
After training embedding models or using some information like category, price, value etc. to represent entities, it is natural to utilize search strategies to select suitable items to form bundles, or using generative models to generate bundles automatically. As shown in Table \ref{generative:bundle generation}, we focus on two lines about bundle generation: search-based and model-based bundle generation, respectively.

\begin{table}[htbp]
\vspace{-0.3cm}
\captionsetup{justification=centering}
\caption{Bundle Generation Methods}
\vspace{-0.3cm}
\label{generative:bundle generation}
\centering
\scalebox{0.68}{
\begin{tabular}{ccc|ccc}
\toprule
\textbf{Category}                      & \textbf{Techniques}  & \textbf{References}                                                                                                                                                                                                               & \textbf{Category}                     & \textbf{Techniques}                                                                            & \textbf{References}                                                                                                                                                                                 \\ \hline
\multirow{3}{*}{\textbf{Search-based}} & Greedy Search        & \begin{tabular}[c]{@{}c@{}}\cite{fang2014trip, CSBR, BundleBPR, GRAM-SMOT, TRAR}\\ \cite{yu2015personalized, benouaret2017recommending, benouaret2016package, xie2010breaking, ma2024cirp}\end{tabular} & \multirow{3}{*}{\textbf{Model-based}} & \begin{tabular}[c]{@{}c@{}}Sequential Generative Models\\ (RNNs, Bi-LSTMs et al.)\end{tabular} & \begin{tabular}[c]{@{}c@{}}\cite{zheng2023interaction, BGN, han2017learning, CATHI, DeepTrip}\\ \cite{li2017mining, yu2023basket, zhu2023text2bundle, CTLTR}\end{tabular} \\ \cline{2-3} \cline{5-6} 
                                       & Beam Search          & \cite{BGN, BGGN, yu2015personalized, zhang2024encoder, sun2023generative}                                                                                                                                            &                                       & \begin{tabular}[c]{@{}c@{}}Adversarial Generative Models\\ (GANs et al.)\end{tabular}          & \cite{liu2021dysr, li2023next}                                                                                                                                                         \\ \cline{2-3} \cline{5-6} 
                                       & Breadth-First Search & \cite{API-Prefer}                                                                                                                                                                                                    &                                       & \begin{tabular}[c]{@{}c@{}}Seq2Seq Generative Models\\ (encoder-decoder et al.)\end{tabular}   & \begin{tabular}[c]{@{}c@{}}\cite{wei2022towards, yang2024nonautoregressive, POG}\\ \cite{sun2023generative, wang2023query2trip}\end{tabular}                              \\ \bottomrule
\end{tabular}
}
\vspace{-0.5cm}
\end{table}

\subsubsection{Search-based}
One line of search-based bundle generation is leveraging greedy algorithm, which is a simple and efficient heuristic search strategy. Its basic idea is to select the local optimal solution at each step and gradually construct the global optimal solution. As for bundle generation, greedy search begins with an empty bundle and iteratively builds the bundle by selecting items one at a time. At each step, it evaluates all possible items that could be added to the current bundle using a scoring function that considers factors like user preferences and item compatibility. The item with the highest score is added to the bundle. This process repeats until the bundle reaches the desired size or another stopping criterion is met, always choosing the locally optimal item to add at each step to create the final bundle. There are many research \cite{xie2010breaking, API-PROGRAM, CSBR, BundleBPR, ma2024cirp} on selecting the optimal items through greedy strategy to form bundles that can maximizing the semantic similarity, utility function, or revenue. A heuristic algorithm is used in \cite{benouaret2017recommending, benouaret2016package, yu2015personalized}, which constructs bundles by greedily selecting the next item that maximizes the bundle score. BundleBPR \cite{BundleBPR} uses a greedy search to find the bundle with the highest score, where the bundle score is computed by Bundle BPR model leveraging the parameters learned from item BPR model. GRAM-SMOT \cite{GRAM-SMOT} framework is proposed, which includes a greedy strategy based on the graph attention mechanism for generating new personalized bundles by maximizing the monotone submodular function. The greedy search has effective computation. However, it may miss the global optimal solution.

Another line is using beam search to find suitable items. Beam search starts with an empty bundle and a list of candidate bundles, or \textquotesingle beam\textquotesingle, containing just the empty bundle. At each step, new candidate bundles are created by adding one item from the item sets to each bundle in the beam. These bundles are scored, and the top K are selected to form the new beam. This iterative process continues until the bundles reach the desired size or meeting stopping criteria, balancing breadth and depth to find the optimal bundle. There are many research \cite{BGN, BGGN} leveraging beam strategy to form bundles. BGN \cite{BGN} is proposed to address the personalized bundle list recommendation problem. It decomposes the problem into quality and diversity parts using Determinantal Point Processes (DPPs). And a masked beam search along with DPP selection is integrated to produce high-quality and diversified bundle lists. In the BGGN model \cite{BGGN}, a diverse beam search method is proposed for decoding diverse bundles on a graph. It starts with a set of candidate items and iteratively chooses the most relevant ones based on learned embeddings and graph density, ensuring both accuracy and diversity in the generated bundles. The process stops when the graph density drop rate falls below a certain threshold, resulting in a personalized bundle of items.

Greedy search and beam search have also been employed to construct personalized travel routes \cite{TRAR, fang2014trip, yu2015personalized, zhang2024encoder}. TRAR \cite{TRAR} uses greedy search to generate candidate set by gradually adding POIs to build the optimal travel package that satisfy multiple user-specific constraints. Yu et al. \cite{yu2015personalized} utilize a path-tree based beam search method, focusing on iteratively selecting POIs based on the highest transition probabilities to form travel routes. In contrast, the work by Zhang et al. \cite{zhang2024encoder} integrates user preferences, geographic information, and a grid beam search approach within an iterative process to select candidate POIs, generating travel routes that meet individualized needs and constraints. Furthermore, dynamic programming and integer programming have been used to solve search problem of forming optimal travel packages \cite{RANDOM, lim2015personalized}.

\subsubsection{Model-based}
Search-based methods generate bundles from the candidate set by searching for the optimal combinations. These methods can explore different combinations more intuitively, but face two problems: first, the search process often relies on predefined heuristics, so the generation effect may not be flexible enough in complex scenarios. Second, the search space expands drastically with the increase of the candidate set, leading to a significant increase in computational overhead. At this point, model-based can cope with these problems and better capture complex relationships, generating bundles with higher quality. As introduced in Section \ref{generative:representation learning}, embedding models and feature extraction techniques are used for distilling features, representing data, and learning compatibility between items. The subsequent phase involves the utilization of these encoded representations by generative models, such as RNNs and GANs. These models are tasked with the generation of potential item bundles that are contextually relevant and aligned with user preferences.

This comprehensive approach to bundle generation is exemplified by models like \cite{yang2024nonautoregressive, wei2022towards}, which introduce an encoder-decoder framework capable of generating personalized bundles in a non-autoregressive manner. In contrast, GenRec \cite{sun2023generative} generates the recommended items one by one to form the next bundle via an autoregressive decoder. DPG \cite{zheng2023interaction} integrates a RNN-based generative model enhanced by a mask layer sensitive to patient conditions, trained via a hybrid approach of maximum likelihood estimation and reinforcement learning to minimize order dependency. DySR \cite{liu2021dysr}, on the other hand, leverages a GAN-style training strategy to dynamically generate service packages, focusing on the evolution of service representations in a shared semantic space, which helps address cold start challenges and align services with user requirements.

In the fashion domain, methods like those proposed by Li et al. \cite{li2017mining} and Han et al. \cite{han2017learning} have traditionally treated fashion bundles as sets or ordered sequences, relying on models such as bidirectional LSTMs for generation. These methods rely heavily on the order of the outfit items. But actually, it is unreasonable to regard an outfit as an ordered sequence, because shuffling items in the outfit should make no difference on its compatibility. Recognizing the limitations of these approaches, Chen et al. \cite{POG} introduced POG, an encoder-decoder model that harnesses the power of Transformer architecture to connect user preferences across individual items and outfits, thereby personalizing the fashion outfit generation process. Similar to the approach taken by POG, DiFashion \cite{xu2024diffusion} leverages the Transformer architecture's ability to model the relationships between items holistically, and generates outfits by parallel conditional denoising, considering the compatibility and personalization without being constrained by item order.

In trip recommendation, He et al. \cite{he2019joint} presents C-ILP and C-ALNS based on a joint context-aware embedding. Besides, PT-VPMF \cite{zhao2021photo2trip} uses CNNs to capture visual contents in geo-tagged photos, then introduces visual-enhanced probabilistic matrix factorization for personalized recommendation. Due to the outstanding performance of the encoder-decoder model in the TripRec, other methods such as CATHI \cite{CATHI}, DeepTrip \cite{DeepTrip}, and CTLTR \cite{CTLTR} follow a similar line of reasoning by also employing the RNN or LSTM based model to learn underlying semantic relationships between POIs in the trajectory. CTLTR \cite{CTLTR} captures semantic sub-trajectories and corresponding sequential information by using RNN and contrastive learning. Different from these methods, Query2Trip\cite{wang2023query2trip} combines sequence generation with Transformer architecture and debiased learning in TripRec modeling to learn more effective representations.

These models collectively represent the state-of-the-art in generative modeling for bundle generation, showcasing an intricate interplay between encoding, generation, and decoding processes that are finely tuned to the nuances of user preferences and item characteristics.

\subsection{Summary of Existing Generative BR Research}
As presented in Table \ref{summary of GBR}, we systematically summarize and categorize the existing Generative BR research work in terms of the dimensions of representation learning and bundle generation. Table \ref{summary of GBR} lists different representation learning inputs ranging from user-bundle, user-item, and bundle-item associations to considering auxiliary information such as visual and textual information about item, cost, budget, and other factors. Meanwhile, generative directions for research include new bundles, new travel packages, new service packages, etc., with each category accompanied by corresponding literature citations. This summary not only echoes the previous two subsections, but also provides a overview and categorical summary of existing Generative BR research.
\begin{table}[htbp]
\vspace{-0.5cm}
\captionsetup{justification=centering}
\caption{Summary of generative bundle recommendation research}
\vspace{-0.3cm}
\label{summary of GBR}			
\centering
\scalebox{0.66}{
\begin{tabular}{ccc}
\toprule
\textbf{Learning}                                                                                                                                 & \textbf{Generation}         & \textbf{References}                                                                                                                                                        \\ \hline
User-Bundle, User-Item                                                                                                                            & \multirow{5}{*}{New Bundle} & \cite{BundleBPR}                                                                                                                                                           \\
User-Bundle, Bundle-Item                                                                                                                          &                             & \cite{BGGN}                                                                                                                                                                \\
User-Bundle, User-Item, Bundle-Item                                                                                                               &                             & \cite{GRAM-SMOT}                                                                                                                                                           \\
\begin{tabular}[c]{@{}c@{}}User-Item, Side information\\ (visual and textual information of items, cost, budget, et al.)\end{tabular}             &                             & \cite{POG, yang2024nonautoregressive, interdonato2013versatile, beladev2016recommender}                                                                                    \\
\begin{tabular}[c]{@{}c@{}}User-Bundle, User-Item\\ Side information (features, context information et al.)\end{tabular}                          &                             & \cite{liu2017modeling, wei2022towards}                                                                                                                                     \\ \hline
\begin{tabular}[c]{@{}c@{}}Users, POIs, Side information\\ (categories of POIs, physical constraints,\\ requirements of users, et al.)\end{tabular} & New Travel Package          & \begin{tabular}[c]{@{}c@{}}\cite{chen2015package, benouaret2016package, xie2010breaking,TRAR}\\ \cite{yu2015personalized, benouaret2017recommending, fang2014trip,RANDOM}\end{tabular} \\ \hline
\begin{tabular}[c]{@{}c@{}}Services, Mashups\\ Service relations, Mashup descriptions\end{tabular}                                                & New Service Package         & \begin{tabular}[c]{@{}c@{}}\cite{CSBR, liu2023dysr, liu2021dysr, API-Prefer}\\ \cite{API-PROGRAM, gu2016service, cao2016service}\end{tabular}                              \\ \hline
Patient descriptions, Drug relations                                                                                                              & New Drug Package            & \cite{zheng2023interaction, zhu2024medicine}                                                                                                                               \\ \hline
\begin{tabular}[c]{@{}c@{}}Bundle list, Side information \\ (user context, et al.)\end{tabular}                                                   & New Bundle List             & \cite{BGN}                                                                                                                                                                 \\ \bottomrule
\end{tabular}
}
\vspace{-0.5cm}
\end{table}

\section{Resources of Bundle Recommendation Research}
\label{section5:applications of BR}
This section first reports the resources of bundle recommendation including datasets from various application domains and evaluation metrics. And we conduct reproducibility experiments on discriminative and generative bundle recommendation models based on the resources.

\subsection{Datasets in Various Domains}
We summarize 30 datasets used for bundle recommendation research across 11 application domains. These datasets are listed in Table \ref{datasets} and categorized by domain including music playlist, book list, movie list, outfit, meal, drug package, service package and so on. Table \ref{datasets} illustrates the domains, sources, the type of data each dataset contains, and how the bundle in each dataset is presented. Moreover, we also present whether content information, such as text, images, or acoustics, is available in datasets, which provides researchers with valuable entry points for leveraging LLMs to further explore bundle recommendation.

\begin{table}[h]
\vspace{-0.3cm}
\caption{Datasets of Bundle Recommendation}
\vspace{-0.3cm}
\label{datasets}
\centering
\resizebox{\textwidth}{!}{
\begin{tabular}{llllllcl}
\hline
\textbf{Domain}                                                                           & \textbf{Name of Dataset} & \textbf{Source}                                                                                                                                & \textbf{Data}                                                                                                                                                                                                             & {\textbf{Content Information}}                                                                                                                                          & \textbf{Description of Bundle}                                                                                                                                    & \textbf{Available or Not} & \textbf{References}                                                                                                                          \\ \hline
                                                                                          & \textbf{NetEase}         & \textbf{Netease Cloud Music}                                                                                                                   & \textbf{\begin{tabular}[c]{@{}l@{}}users,songs,playlist,\\ user-song,user-list\end{tabular}}                                                                                                                              & { \textbf{-}}                                                                                                                                                & \textbf{\begin{tabular}[c]{@{}l@{}}A playlist contains user-defined \\ collections of songs on a specific theme.\end{tabular}}                                    & \textbf{Y}                & \textbf{\begin{tabular}[c]{@{}l@{}}\cite{BGCN,MIDGN}\\ \cite{EFM,CrossCBR}\end{tabular}}                                                                   \\ \cline{2-8} 
                                                                                          & \textbf{Spotify}         & \textbf{Spotify}                                                                                                                               & { \textbf{\begin{tabular}[c]{@{}l@{}}users,items,lists,\\ user-list,list-item,density\end{tabular}}}                                                                                                  & { \textbf{\begin{tabular}[c]{@{}l@{}}music titles, playlist title,\\ album name and artist names\\ Spotify URIs and the playlist’s \\ number of followers\end{tabular}}} & \textbf{A playlist is composed of songs with popular topics.}                                                                                                     & \textbf{Y}                & \textbf{\cite{AttList,he2020consistency}}                                                                                                           \\ \cline{2-8} 
\multirow{-3}{*}{\textbf{Playlist}}                                                       & \textbf{AOTM}            & \textbf{Art of the Mix}                                                                                                                        & \textbf{\begin{tabular}[c]{@{}l@{}}users,items,lists,\\ list-item,\\ avg. list interactions,\\ avg. list size,density\end{tabular}}                                                                                       & { \textbf{\begin{tabular}[c]{@{}l@{}}category, Mixed Genre, \\ timestamp, etc.\end{tabular}}}                                                                            & \textbf{A playlist is composed of songs.}                                                                                                                         & \textbf{Y}                & \textbf{\cite{mcfee2012hypergraph,he2020consistency}}                                                                                               \\ \hline
\textbf{Booklist}                                                                         & \textbf{Youshu}          & \textbf{a book review site Youshu}                                                                                                             & \textbf{\begin{tabular}[c]{@{}l@{}}users,books,lists,\\ user-list,user-book,list-book,\\ avg. list interactions,\\ avg. book interactions,\\ avg. list size,density\end{tabular}}                                         & { \textbf{-}}                                                                                                                                                & \textbf{A booklist is a list of books under different genres.}                                                                                                    & \textbf{Y}                & \textbf{\begin{tabular}[c]{@{}l@{}}\cite{DAM,CrossCBR}\\ \cite{BundleGT,DGMAE}\end{tabular}}                                                               \\ \cline{2-8} 
                                                                                          & \textbf{Goodreads}       & \textbf{\begin{tabular}[c]{@{}l@{}}a popular site for book \\ reviews and recommendation\end{tabular}}                                         & \textbf{\begin{tabular}[c]{@{}l@{}}users,books,lists,\\ user-book,user-list,\\ avg. book interactions\end{tabular}}                                                                                                       & { \textbf{\begin{tabular}[c]{@{}l@{}}book title, language, \\ description, genre, \\ authors, etc.\end{tabular}}}                                                        & \textbf{A booklist is a list of books.}                                                                                                                           & \textbf{Y}                & \textbf{\begin{tabular}[c]{@{}l@{}}\cite{AttList,he2020consistency}\\ \cite{LIRE}\end{tabular}}                                                            \\ \hline
                                                                                          & \textbf{MovieLens}       & \textbf{\begin{tabular}[c]{@{}l@{}}published by GroupLens \\ research group\end{tabular}}                                                      & \textbf{\begin{tabular}[c]{@{}l@{}}users,movies,\\ user-movie ratings\end{tabular}}                                                                                                                                       & { \textbf{\begin{tabular}[c]{@{}l@{}}user gender, age, occupation,\\ movie titles, genres\end{tabular}}}                                                                 & \textbf{-}                                                                                                                                                        & \textbf{Y}                & \textbf{\begin{tabular}[c]{@{}l@{}}\cite{xie2010breaking,BundleMCR}\\ \cite{kabutoya2013probabilistic}\end{tabular}}                                       \\ \cline{2-8} 
                                                                                          & \textbf{Douban}          & \textbf{Douban}                                                                                                                                & \textbf{\begin{tabular}[c]{@{}l@{}}users,items,lists,\\ user-item,user-list\end{tabular}}                                                                                                                                 & { \textbf{-}}                                                                                                                                                & \textbf{A list is movie watching list.}                                                                                                                           & \textbf{Y}                & \textbf{\cite{GANCF}}                                                                                                                               \\ \cline{2-8} 
\multirow{-3}{*}{\textbf{Movie}}                                                          & \textbf{MX Player}       & \textbf{\begin{tabular}[c]{@{}l@{}}MX Player, one of the largest \\ streaming services in India\end{tabular}}                                  & \textbf{\begin{tabular}[c]{@{}l@{}}users,lists,items,\\ user-list,user-item,\\ density\end{tabular}}                                                                                                                      & { \textbf{\begin{tabular}[c]{@{}l@{}}list names,\\ list contents\end{tabular}}}                                                                                          & \textbf{A list is a set of items with different contents.}                                                                                                        & \textbf{N}                & \textbf{\cite{CAPLE,MULTIPLE}}                                                                                                                      \\ \hline
                                                                                          & \textbf{iFashion}        & { \textbf{\begin{tabular}[c]{@{}l@{}}originating from \\ the POG dataset\end{tabular}}}                                    & { \textbf{\begin{tabular}[c]{@{}l@{}}users,items,outfits,\\ user-outfit,user-item\end{tabular}}}                                                                                                      & { \textbf{\begin{tabular}[c]{@{}l@{}}fashion item category, \\ images, title, etc.\end{tabular}}}                                                                        & \textbf{An outfit is a set of well-matched fashion items.}                                                                                                        & \textbf{Y}                & \textbf{\begin{tabular}[c]{@{}l@{}}\cite{POG,CrossCBR}\\ \cite{EBRec,MultiCBR}\end{tabular}}                                                               \\ \cline{2-8} 
                                                                                          & \textbf{Clothing}        & { \textbf{originating from Amazon}}                                                                                        & { \textbf{\begin{tabular}[c]{@{}l@{}}users,items,bundles,\\ user-bundle,user-item\end{tabular}}}                                                                                                      & { \textbf{bundle intents, category, title}}                                                                                                                              & \textbf{A bundle consists of several clothes.}                                                                                                                    & \textbf{Y}                & \textbf{\cite{sun2022revisiting,BundleMCR}}                                                                                                         \\ \cline{2-8} 
                                                                                          & \textbf{Taobao}          & \textbf{\begin{tabular}[c]{@{}l@{}}from an online bundle \\ list recommendation \\ service of a commercial \\ e-business, Taobao\end{tabular}} & \textbf{\begin{tabular}[c]{@{}l@{}}users,items,bundles,\\ categories,records,\\ avg. bundle size\end{tabular}}                                                                                                            & { \textbf{-}}                                                                                                                                                & \textbf{A bundle is predefined manually by the sellers.}                                                                                                          & \textbf{N}                & \textbf{\cite{BGN}}                                                                                                                                 \\ \cline{2-8} 
\multirow{-4}{*}{\textbf{Outfit}}                                                         & \textbf{Polyvore}        & \textbf{\begin{tabular}[c]{@{}l@{}}Polyvore, a well-known \\ fashion website\end{tabular}}                                                     & { \textbf{\begin{tabular}[c]{@{}l@{}}users,items,outfits,\\ user-outfit\end{tabular}}}                                                                                                                & { \textbf{\begin{tabular}[c]{@{}l@{}}fashion item descriptions, images,\\ number of likes of the outfit, \\ hash tags of the outfits, etc.\end{tabular}}}                & \textbf{\begin{tabular}[c]{@{}l@{}}An outfit contains multiple clothing items,organizing \\ in fixed order-tops,bottoms,shoes,and accessories.\end{tabular}}      & \textbf{Y}                & \textbf{\cite{han2017learning}}                                                                                                                     \\ \hline
                                                                                          & \textbf{Steam}           & \textbf{\begin{tabular}[c]{@{}l@{}}Steam video game \\ distribution platform\end{tabular}}                                                     & \textbf{\begin{tabular}[c]{@{}l@{}}users,games,bundles,\\ user-game,user-bundle,\\ avg. bundle size\end{tabular}}                                                                                                         & { \textbf{\begin{tabular}[c]{@{}l@{}}bundle name, bundle price, \\ bundle url, genre, item name, \\ item url, pricing information\end{tabular}}}                         & \textbf{\begin{tabular}[c]{@{}l@{}}A bundle is a group of different \\ games sold together on website.\end{tabular}}                                              & \textbf{Y}                & \textbf{\begin{tabular}[c]{@{}l@{}}\cite{BundleBPR,BundleNet}\\ \cite{BRUCE,PopCon}\end{tabular}}                                                          \\ \cline{2-8} 
                                                                                          & \textbf{Justice}         & \textbf{\begin{tabular}[c]{@{}l@{}}the mobile game \\ Love is Justice\\ developed by \\ Netease Games\end{tabular}}                            & \textbf{\begin{tabular}[c]{@{}l@{}}users,items,bundles,\\ user-bundle,user-item,bundle-item\end{tabular}}                                                                                                                 & { \textbf{-}}                                                                                                                                                & \textbf{\begin{tabular}[c]{@{}l@{}}A bundle is made up of props\\ (virtual items) in the game.\end{tabular}}                                                      & \textbf{-}                & \textbf{\cite{BundleNet}}                                                                                                                           \\ \cline{2-8} 
\multirow{-3}{*}{\textbf{Game List}}                                                      & \textbf{Games}           & \textbf{\begin{tabular}[c]{@{}l@{}}NetEase Games’ \\ massively multiplayer \\ online role-playing game\end{tabular}}                           & \textbf{\begin{tabular}[c]{@{}l@{}}users,items,bundles,\\ avg. bundle interactions,\\ avg. item interactions,\\ avg. bundle size\end{tabular}}                                                                            & { \textbf{-}}                                                                                                                                                & \textbf{\begin{tabular}[c]{@{}l@{}}A bundle consists of the virtual \\ items sold in the game.\end{tabular}}                                                      & \textbf{-}                & \textbf{\cite{AGF}}                                                                                                                                 \\ \hline
\textbf{Meal}                                                                             & \textbf{MealRec}         & \textbf{\begin{tabular}[c]{@{}l@{}}Allrecipes, one popular \\ food sharing site in the world\end{tabular}}                                     & { \textbf{\begin{tabular}[c]{@{}l@{}}users,recipes,meals,\\ user-meal,user-recipe,meal-recipe\end{tabular}}}                                                                                          & { \textbf{\begin{tabular}[c]{@{}l@{}}recipe name, image, category,\\  rate, ingredients, nutrition, tags\end{tabular}}}                                                  & \textbf{\begin{tabular}[c]{@{}l@{}}A meal is a bundle composed of \\ three recipes from different categories.\end{tabular}}                                       & \textbf{Y}                & \textbf{\cite{MealRec,li2024mealrec+,li2023user}}                                                                                                   \\ \hline
                                                                                          & \textbf{Tuniu}           & \textbf{\begin{tabular}[c]{@{}l@{}}one of the largest tourism \\ e-commerce platforms in China\end{tabular}}                                   & \textbf{\begin{tabular}[c]{@{}l@{}}users,items,records,\\ sessions,purchased items\end{tabular}}                                                                                                                          & { \textbf{-}}                                                                                                                                                & \textbf{\begin{tabular}[c]{@{}l@{}}A package contains a set of travel-related attributes,\\ such as title, destination, travel region, travel type.\end{tabular}} & \textbf{Y}                & \textbf{\cite{STR-VGAE,NATR}}                                                                                                                       \\ \cline{2-8} 
                                                                                          & \textbf{TripAdvisor}     & \textbf{a well-known travel website}                                                                                                           & \textbf{\begin{tabular}[c]{@{}l@{}}users,POIs,\\ user-POI ratings\end{tabular}}                                                                                                                                           & { \textbf{\begin{tabular}[c]{@{}l@{}}POI thematic category, \\ the average rating score, \\ the number of ratings, etc.\end{tabular}}}                                   & \textbf{\begin{tabular}[c]{@{}l@{}}A package is constituted with a set of different POIs\\ that may constitute a tour.\end{tabular}}                              & \textbf{-}                & \textbf{\begin{tabular}[c]{@{}l@{}}\cite{benouaret2016package, interdonato2013versatile}\\ \cite{xie2010breaking, benouaret2017recommending}\end{tabular}} \\ \cline{2-8} 
                                                                                          & \textbf{Flickr}          & \textbf{\begin{tabular}[c]{@{}l@{}}the largest public multimedia \\ collection released on Flickr\end{tabular}}                                & \textbf{\begin{tabular}[c]{@{}l@{}}users,POIs,trajectories,\\ user-POI visits\end{tabular}}                                                                                                                               & { \textbf{\begin{tabular}[c]{@{}l@{}}POI photo, longitude, latitude, theme, \\ dateTaken, frequency, travel sequence\end{tabular}}}                                      & \textbf{A trajectory is composed of several ordered POIs.}                                                                                                        & \textbf{Y}                & \textbf{\begin{tabular}[c]{@{}l@{}}\cite{RANDOM,wang2023query2trip}\\ \cite{he2019joint,gao2022selfsupervised}\end{tabular}}                               \\ \cline{2-8} 
                                                                                          & \textbf{Foursquare}      & \textbf{\begin{tabular}[c]{@{}l@{}}a geo-social networking dataset \\ published by Foursquare\end{tabular}}                                    & \textbf{\begin{tabular}[c]{@{}l@{}}users,POIs,trajectories,\\ user-POI visits\end{tabular}}                                                                                                                               & { \textbf{\begin{tabular}[c]{@{}l@{}}time stamp, GPS coordinates,\\ venue-categories, tag\end{tabular}}}                                                                 & \textbf{A trajectory is composed of several ordered POIs.}                                                                                                        & \textbf{Y}                & \textbf{\begin{tabular}[c]{@{}l@{}}\cite{RANDOM,wang2023query2trip}\\ \cite{he2019joint,gao2022selfsupervised}\end{tabular}}                               \\ \cline{2-8} 
\multirow{-5}{*}{\textbf{Travel Package}}                                                 & \textbf{Gowalla}         & \textbf{\begin{tabular}[c]{@{}l@{}}a real check-in dataset \\ obtained from Gowalla\end{tabular}}                                              & { \textbf{\begin{tabular}[c]{@{}l@{}}users,POIs,packages\\ user-POI check-ins,\\ user-pacakge check-ins,\\ package-POI,\\ avg. POI check-in per user,\\ avg. package check-in per user\end{tabular}}} & { \textbf{\begin{tabular}[c]{@{}l@{}}POI category, stay time, cost,\\ package total cost, total stay time,\\ distances between POIs,\\ user features\end{tabular}}}      & \textbf{A package is orderly composed of several POIs.}                                                                                                           & \textbf{Y}                & \textbf{\cite{fang2014trip}}                                                                                                                        \\ \hline
                                                                                          & \textbf{EMR}             & \textbf{\begin{tabular}[c]{@{}l@{}}from the electronic \\ medical record database \\ of a first-rate hospital in China\end{tabular}}           & { \textbf{\begin{tabular}[c]{@{}l@{}}patients, visit records, demographics, \\ labs, prescriptions, drug packages, \\ explicit drug–drug interactions\end{tabular}}}                                  & { \textbf{\begin{tabular}[c]{@{}l@{}}admission notes, \\ disease description documents\end{tabular}}}                                                                    & \textbf{\begin{tabular}[c]{@{}l@{}}A package consists of several drugs with drug interactions, \\ such as synergism, antagonism or no interaction.\end{tabular}}  & \textbf{N}                & \textbf{\cite{zheng2023interaction}}                                                                                                                \\ \cline{2-8} 
\multirow{-2}{*}{\textbf{Drug Package}}                                                   & \textbf{MIMIC-III}       & \textbf{\begin{tabular}[c]{@{}l@{}}a freely accessible critical \\ medical database\end{tabular}}                                              & { \textbf{\begin{tabular}[c]{@{}l@{}}patients, admissions, ICU stays, \\ vitals, labs, medications, procedures, \\ diagnostic codes, survival outcomes\end{tabular}}}                                 & { \textbf{\begin{tabular}[c]{@{}l@{}}clinical notes, discharge summaries, \\ imaging reports,\\ physiological waveforms\end{tabular}}}                                   & \textbf{\begin{tabular}[c]{@{}l@{}}A package consists of several drugs with drug interactions, \\ such as synergism, antagonism or no interaction.\end{tabular}}  & \textbf{Y}                & \textbf{\cite{zheng2023interaction,johnson2016mimic}}                                                                                               \\ \hline
\textbf{Service Package}                                                                  & \textbf{ProgrammableWeb} & \textbf{\begin{tabular}[c]{@{}l@{}}the largest mashup information \\ sharing community\end{tabular}}                                           & { \textbf{\begin{tabular}[c]{@{}l@{}}mashups,services,\\ mashup-service,\\ avg. service per mashup\end{tabular}}}                                                                                     & { \textbf{\begin{tabular}[c]{@{}l@{}}functional description, name, \\ url, category, format, provider, \\ developer, etc.\end{tabular}}}                                 & \textbf{\begin{tabular}[c]{@{}l@{}}A package composes of compatible services\\ satisfying the functional requirements of the mashup.\end{tabular}}                & \textbf{Y}                & \textbf{\begin{tabular}[c]{@{}l@{}}\cite{CSBR,liu2023dysr}\\ \cite{API-Prefer,API-PROGRAM}\end{tabular}}                                                   \\ \hline
\textbf{QA List}                                                                          & \textbf{Zhihu}           & \textbf{\begin{tabular}[c]{@{}l@{}}a question and \\ answer community\end{tabular}}                                                            & { \textbf{\begin{tabular}[c]{@{}l@{}}users,answers,lists,\\ user-list,list-answer\end{tabular}}}                                                                                                      & { \textbf{\begin{tabular}[c]{@{}l@{}}answer content, question semantics,\\ list semantics, topics\end{tabular}}}                                                         & \textbf{A list contains several answers of popular topics.}                                                                                                       & \textbf{Y}                & \textbf{\cite{AttList}}                                                                                                                             \\ \hline
                                                                                          & \textbf{cosmestics}      & \textbf{\begin{tabular}[c]{@{}l@{}}from one online \\ merchant in Taobao\end{tabular}}                                                         & { \textbf{\begin{tabular}[c]{@{}l@{}}customers,skin care products,\\ user-item transaction records,\\ user-bundle transaction records\end{tabular}}}                                                  & { \textbf{\begin{tabular}[c]{@{}l@{}}product name, brand, price, \\ functionality, origin, fragrance,\\ transaction price variance, \\ price distribution\end{tabular}}} & \textbf{\begin{tabular}[c]{@{}l@{}}A product bundle consists of two or more items transacted \\ by users.\end{tabular}}                                           & \textbf{-}                & \textbf{\cite{liu2017modeling}}                                                                                                                     \\ \cline{2-8} 
                                                                                          & \textbf{perfume}         & \textbf{\begin{tabular}[c]{@{}l@{}}from one online \\ merchant in Taobao\end{tabular}}                                                         & { \textbf{\begin{tabular}[c]{@{}l@{}}customers,perfumes,\\ user-item transaction records, \\ user-bundle transaction records\end{tabular}}}                                                           & { \textbf{\begin{tabular}[c]{@{}l@{}}product name, brand, price, \\ functionality, origin, fragrance,\\ transaction price variance, \\ price distribution\end{tabular}}} & \textbf{\begin{tabular}[c]{@{}l@{}}A product bundle consists of two or more items transacted \\ by users.\end{tabular}}                                           & \textbf{-}                & \textbf{\cite{liu2017modeling}}                                                                                                                     \\ \cline{2-8} 
                                                                                          & \textbf{TaFeng}          & \textbf{\begin{tabular}[c]{@{}l@{}}a Chinese grocery\\ store dataset\end{tabular}}                                                             & \textbf{\begin{tabular}[c]{@{}l@{}}users,items,baskets,\\ interactions,\\ avg.basket size\end{tabular}}                                                                                                                   & { \textbf{\begin{tabular}[c]{@{}l@{}}transaction date and time, \\ age group, PIN code, product subclass,\\ amount, asset, sales price\end{tabular}}}                    & \textbf{A basket consists of the items bought in the same order.}                                                                                                 & \textbf{Y}                & \textbf{\begin{tabular}[c]{@{}l@{}}\cite{yu2023basket,deng2023multiview}\\ \cite{sun2023generative,su2024hierarchical}\end{tabular}}                       \\ \cline{2-8} 
                                                                                          & \textbf{Dunnhumby}       & \textbf{\begin{tabular}[c]{@{}l@{}}a household-level \\ transaction dataset\end{tabular}}                                                      & \textbf{\begin{tabular}[c]{@{}l@{}}users,items,baskets,\\ interactions,\\ avg.basket size\end{tabular}}                                                                                                                   & { \textbf{\begin{tabular}[c]{@{}l@{}}description for the campaign, \\ marital status, income description, \\ product department, brand, \\ customer age\end{tabular}}}   & \textbf{A basket consists of the items bought in the same order.}                                                                                                 & \textbf{Y}                & \textbf{\begin{tabular}[c]{@{}l@{}}\cite{yu2023basket,deng2023multiview}\\ \cite{li2023next,deng2023multiaspect}\end{tabular}}                             \\ \cline{2-8} 
\multirow{-5}{*}{\textbf{\begin{tabular}[c]{@{}l@{}}Co-consumed\\ Products\end{tabular}}} & \textbf{Instacart}       & \textbf{\begin{tabular}[c]{@{}l@{}}an online grocery \\ shopping dataset \\ published by instacart\end{tabular}}                               & \textbf{\begin{tabular}[c]{@{}l@{}}users,items,baskets,\\ interactions,\\ avg.basket size\end{tabular}}                                                                                                                   & { \textbf{\begin{tabular}[c]{@{}l@{}}aisle category,\\ department category, \\ item name\end{tabular}}}                                                                  & \textbf{\begin{tabular}[c]{@{}l@{}}A basket is constructed with the user's grocery\\ transaction records from online grocery shopping.\end{tabular}}              & \textbf{Y}                & \textbf{\cite{yu2023basket,BasConv}}                                                                                                                \\ \hline
\end{tabular}
}
\vspace{-0.5cm}
\end{table}

\textbf{Music.} A playlist usually consists of songs on specific theme. The typical one is NetEase constructed by \cite{EFM}. They have collected data from Netease Cloud Music, which enables users to construct a list of songs with a specific theme. Another commonly utilized dataset is Spotify dataset \cite{AttList, he2020consistency}. Although the Spotify adopted in these studies are mainly constructed with interaction data, the original Spotify dataset actually contains richer content information, such as playlist titles, artist names, URLs, etc. Additionally, AOTM dataset, proposed by \cite{mcfee2012hypergraph}, was created by collecting playlists from Art of the Mix, a website where users can upload their song playlists.

\textbf{Book.} A booklist contains multiple books with different genres. One prominent dataset in this domain is named Youshu, provided by \cite{DAM}, which is constructed by crawling data from Youshu, a Chinese book review site. Similarly, Goodreads dataset which serves as a comprehensive repository of book-related information has been leveraged in studies \cite{AttList, he2020consistency, LIRE}, comprising book title, language, description, genre, user-generated reviews, ratings, etc.

\textbf{Movie.} A movie list is a set of movies. One notable dataset in this domain is the MovieLens, collected rating data sets from the MovieLens web site by GroupLens research group. Since there are only interactions with individual movies in the dataset, bundle recommendation studies usually package movies into lists and then recommend lists to users \cite{xie2010breaking, BundleMCR, kabutoya2013probabilistic}. Another dataset is Douban, crawled from an online social platform Douban by \cite{GANCF}, which enables users to select interested movies or user generated movie watching lists. MX Player, one of the largest streaming platforms in India, organizes and exposes video via lists that include names and contents. Several studies \cite{CAPLE, MULTIPLE} have constructed their own datasets from MX Player to conduct experiments.

\textbf{Outfit.} An outfit is a set of well-matched fashion items, a form of bundle. The widely used iFashion dataset \cite{POG} provides category, image, and title information, and has served as a benchmark in numerous bundle recommendation studies \cite{CrossCBR, DGMAE, EBRec, MultiCBR, CoHEAT, GPCL}. The Clothing dataset \cite{sun2022revisiting}, derived from Amazon, is a public bundle dataset in which each bundle consists of several clothes annotated with intents, categories, and titles. The Polyvore dataset \cite{han2017learning}, collected from the Polyvore website, contains outfits and items with product images, tags, and textual descriptions. The Taobao dataset \cite{BGN}, collected from an online bundle list recommendation service of the e-commerce platform Taobao, contains bundles manually predefined by the sellers.

\textbf{Meal.} A meal contains multiple courses from specific categories. The MealRec dataset, proposed by \cite{MealRec}, is a publicly accessible dataset tailored for meal recommendation research. It is constructed from user review records of Allrecipe.com, including users, recipes and meals. Each recipe is enriched with detailed attributes such as name, rating, image URL, category, tag, and other metadata. Due to the lack of publicly available meal recommendation datasets including meal-course affiliation, MealRec$^+$, a new meal recommendation benchmark dataset, is introduced by \cite{li2024mealrec+}. It fills a data gap by including meal-course affiliation and healthiness information.

\textbf{Game.} A game list is a group of different games sold together on website. The Steam dataset, collected from the Steam video game distribution platform by \cite{BundleBPR}, contains gamers, game bundles, and purchase records for both individual games and bundles. In addition, the Steam dataset provides rich content information such as bundle names, prices, URLs, genres, game titles, and pricing details. The Justice, an industrial dataset, is collected from the mobile game \textit{Love is Justic}e developed by Netease Games, where bundles consist of virtual items in the game. One dataset named Games \cite{AGF} is also collected from NetEase Games, a massively multiplayer online role-playing game, whose bundles consist of the virtual items sold in the game.

\textbf{Travel.} A travel package is an ordered set of multiple POIs, where the order corresponds to the visiting sequence. The Flickr and Foursquare datasets are widely used in \cite{RANDOM, wang2023query2trip, gao2022selfsupervised, he2019joint}. Both provide user–POI interactions along with rich content information including geographical coordinates, timestamps, and categorical attributes. Their differences lie mainly in data sources and emphasis: Flickr collects POIs offering additional features like photos, visit frequencies, and travel sequences, whereas Foursquare records explicit user check-ins enriched with venue tags. The TripAdvisor dataset \cite{xie2010breaking, benouaret2016package, benouaret2017recommending}, derived from the well-known travel website where users can share and explore travel information, provides user reviews, ratings, and check-ins, with each POI further described by thematic categories, average ratings, and rating counts. The Gowalla dataset \cite{fang2014trip} contains users, POIs, and check-ins, along with content information including attraction categories, attraction cost, package compositions, and distances between attractions. Another real-world travel dataset is provided by Tuniu, one of the largest tourism e-commerce platforms in China. It comprises page-to-page clickstream data recorded in server logs and has been adopted in studies \cite{STR-VGAE, NATR}. \looseness=-1

\textbf{Drug.} A drug package refers to a set of drugs prescribed together during a patient’s hospitalization, where drugs may interact through synergism, antagonism, or not at all. Two widely used datasets are MIMIC-III \cite{johnson2016mimic} and EMR \cite{zheng2023interaction}. MIMIC-III provides rich metadata, including patient profiles, admissions, ICU stays, medications, laboratory results, procedure and diagnostic codes, and survival outcomes, as well as content such as clinical notes and imaging reports. In contrast, EMR, constructed from the electronic medical record database of a top hospital in China, emphasizes drug packages with explicit drug–drug interaction annotations, alongside structured records of demographics, lab tests, and prescriptions, as well as admission notes as textual content.

\textbf{Service.} A service package, or API package, is a bundle of compatible services that satisfies the functional requirements of a mashup as fully as possible. The widely used ProgrammableWeb dataset \cite{CSBR, liu2021dysr, liu2023dysr, API-Prefer, API-PROGRAM, gu2016service, cao2016service} is collected from ProgrammableWeb.com, the largest mashup information sharing community. The dataset contains services, mashups, mashup-service relations, and content such as textual descriptions, tags, and categories.

\textbf{Others.} Several other datasets contribute to the field of bundle recommendation research. One such dataset is Zhihu, constructed by \cite{AttList}, which gathers data from Zhihu, a question and answer community where users can follow lists of answers that have been curated by other users. Furthermore, TaFeng, Dunnhumby, and Instacart datasets are widely used for next basket recommendation studies \cite{yu2023basket, sun2023generative, su2024hierarchical, deng2023multiview}. More details can be found in Table \ref{datasets}.

\subsection{Evaluation Metrics}
As shown in Table \ref{tab:metrics}, it illustrates some of the key metrics commonly used in the evaluation of bundle recommendation systems \cite{BGCN, DAM, MIDGN, CrossCBR, MultiCBR, li2024boosting, GANCF}, along with their formulas and brief descriptions. We summarize 4 metrics for accuracy, 3 metrics for ranking quality, 4 metrics for diversity, 2 metrics for ranking exposure, and other metrics like AUC. These metrics provide a comprehensive understanding of how well the system retrieves and ranks relevant bundles. Given that \( \mathbb{I}(\cdot) \) is the indicator function, which is 1 if the condition inside is true, and 0 otherwise; \( \text{threshold} \) is a predefined value used to determine if a prediction is considered positive; \( \text{rank}(b_i) \) is the predicted rank of the bundle \( b_i \) for user \( u \); \( \text{sim}(b_i, b_j) \) is a similarity measure between two bundles \( b_i \) and \( b_j \); $\hat{y}_{ub}$ is the predicted score, and $y_{ub}$ is the actual score; $\mathbf{R}(k)$ is a list of k recommended bundles; $app(i, \mathbf{R}(k))$ indicates whether item $i$ appears in $\mathbf{R}(k)$; $p(i, \mathbf{R}(k))$ indicates the sum of item $i$'s frequency in user $u$'s recommended bundles; $utility(x\mid r)$ denotes the utility function, which is usually a monotonically decreasing function with respect to the rank of a bundle. One common choice is to use an exponent $w>=0$, and $utility(x \mid r)=[\operatorname{rank}(x \mid r)]^{-w}$.
\begin{table}[h]
\vspace{-0.3cm}
\caption{Metrics for evaluating bundle recommendation}
\vspace{-0.3cm}
\label{tab:metrics}
\centering
\resizebox{\textwidth}{!}{
\begin{tabular}{cccl|cccl}
\toprule
\textbf{Category}                                                          & \textbf{Metric} & \textbf{Formulas}                                                                                                                                & \multicolumn{1}{c|}{\textbf{Description}}                                                      & \textbf{Category}                                                           & \textbf{Metric}             & \textbf{Formulas}                                                                                                                                                                           & \multicolumn{1}{c}{\textbf{Description}}                                                                      \\ \hline
\multirow{4}{*}{Accuracy}                                                  & Recall          & $\frac{\sum_{u,b} \mathbb{I}(\hat{y}_{ub} > \text{threshold} \land y_{ub} > 0)}{\sum_{u,b} \mathbb{I}(y_{ub} > 0)}$                              & \begin{tabular}[c]{@{}l@{}}Proportion of\\ relevant bundles\\ recommended.\end{tabular}        & \multirow{4}{*}{Diversity}                                                  & Diversity                   & $1 - \frac{\sum_{i=1}^{N} \sum_{j=1}^{N} \text{sim}(\hat{b}_i, \hat{b}_j)}{N(N-1)}$                                                                                                         & \begin{tabular}[c]{@{}l@{}}Diversity of \\ recommended bundles.\end{tabular}                               \\ \cline{2-4} \cline{6-8} 
                                                                           & HR              & $\frac{\sum_{u,b} \mathbb{I}(\hat{y}_{ub} > \text{threshold} \land y_{ub} > 0)}{\sum_{u,b} \mathbb{I}(y_{ub} > 0)}$                              & \begin{tabular}[c]{@{}l@{}}Proportion of\\ recommended \\ relevant bundles.\end{tabular}       &                                                                             & Coverage                    & $\frac{1}{|\mathcal{I}|} \sum_{i \in \mathcal{I}} \operatorname{app}(i, \mathbf{R}(k))$                                                                                                     & \begin{tabular}[c]{@{}l@{}}Measures how\\ many different\\ items are contained\\ in the results.\end{tabular} \\ \cline{2-4} \cline{6-8} 
                                                                           & Precision       & $\frac{\sum_{u,b} \mathbb{I}(\hat{y}_{ub} > \text{threshold} \land y_{ub} > 0)}{\sum_{u, b} \mathbb{I}(\hat{y}_{ub} > \text{threshold})}$        & \begin{tabular}[c]{@{}l@{}}Proportion of\\ relevant bundles\\ in recommendations.\end{tabular} &                                                                             & Entropy                     & $-\sum_{i \in \mathcal{I}} p(i, \mathbf{R}(k)) \log p(i, \mathbf{R}(k))$                                                                                                                    & \begin{tabular}[c]{@{}l@{}}Measures how\\ evenly all items\\ appear in the results.\end{tabular}              \\ \cline{2-4} \cline{6-8} 
                                                                           & F1              & $2 \times \frac{Precision \times Recall}{Precision + Recall}$                                                                                    & \begin{tabular}[c]{@{}l@{}}Balance between \\ precision and recall.\end{tabular}               &                                                                             & \multicolumn{1}{l}{Novelty} & $\frac{\sum_{i \in R}-\log _{2} p(i)}{|R|}$                                                                                                                                                 & \begin{tabular}[c]{@{}l@{}}Measures the\\ novelty of\\ recommended bundle.\end{tabular}                       \\ \hline
\multirow{3}{*}{\begin{tabular}[c]{@{}c@{}}Ranking\\ Quality\end{tabular}} & NDCG            & $\frac{\sum_{i=1}^{k} \frac{\mathbb{I}(y_{ub_i} > 0)}{\log_2(\text{rank}(b_i) + 2)}}{\sum_{i=1}^{k} \frac{1}{\log_2(\text{ideal\_rank}_i + 2)}}$ & \begin{tabular}[c]{@{}l@{}}Quality of \\ recommended \\ bundle list.\end{tabular}              & \multirow{2}{*}{\begin{tabular}[c]{@{}c@{}}Ranking\\ Exposure\end{tabular}} & Exposure                    & $\sum_{x \in \mathcal{A}} utility(x\mid r)$                                                                                                                                                 & \begin{tabular}[c]{@{}l@{}}Ranking exposure \\ fairness of \\ recommended bundles.\end{tabular}               \\ \cline{2-4} \cline{6-8} 
                                                                           & MAP             & $\frac{1}{N} \sum_{u=1}^{N} AP_u$                                                                                                                & \begin{tabular}[c]{@{}l@{}}Average precision \\ per user.\end{tabular}                         &                                                                             & Gap                         & $\frac{|\operatorname{Exposure}(\mathcal{A} \mid r)-\operatorname{Exposure}(\mathcal{B} \mid r)|}{\operatorname{Exposure}(\mathcal{A} \mid r)+\operatorname{Exposure}(\mathcal{B} \mid r)}$ & \begin{tabular}[c]{@{}l@{}}Ranking exposure\\ gap between two\\ bundle groups.\end{tabular}                   \\ \cline{2-8} 
                                                                           & MRR             & $\frac{1}{N} \sum_{u=1}^{N} \frac{1}{\text{rank}(\hat{y}_{ub^*})}$                                                                               & \begin{tabular}[c]{@{}l@{}}Reciprocal of \\ first relevant \\ bundle's rank.\end{tabular}      & Others                                                                      & AUC                         & Area                                                                                                                                                                                        & \begin{tabular}[c]{@{}l@{}}Distinguishing \\ ability across \\ thresholds.\end{tabular}                       \\ \bottomrule
\end{tabular}
}
\vspace{-0.5cm}
\end{table}

\subsection{Comparison and Experiments}
We conduct reproducibility experiments on bundle recommendation. For discriminative bundle recommendation, we reproduce 6 mainstream graph-based models and 4 CF-based models and compare their performance. For generative bundle recommendation, we focus on trip generation and recommendation, a representative scenario with sufficient resources for empirical experiments.
\subsubsection{Discriminative BR}
\paragraph{\textbf{Experimental Details.}}
We use three real-world public datasets for evaluation, i.e., Youshu, NetEase, and iFashion. Two widely used metrics \cite{BGCN, AttList, MIDGN}, Recall@K and NDCG@K, are adopted to evaluate the top-K recommendation performance, where $K \in \{20, 40\}$.
We reproduce 6 graph-based models, CrossCBR \cite{CrossCBR}, MultiCBR \cite{MultiCBR}, BundleGT \cite{BundleGT}, MIDGN \cite{MIDGN}, BGCN \cite{BGCN}, and HyperMBR \cite{HyperMBR}, 4 CF-based models, LightGCN \cite{LightGCN}, NGCF \cite{NGCF}, ItemKNN \cite{ItemKNN}, MFBPR\cite{rendle2009bpr}, on the Youshu, NetEase and iFashion datasets. Among these, MultiCBR, BundleGT, MIDGN, BGCN and HyperMBR are reproduced on the three datasets using five different random seeds (123, 124, 125, 126, 127) and take the average as the result. For all methods, we follow the optimal parameters provided in the cited papers. We adopt BPR loss, employ Adam to optimize parameters and set the embedding size as 64. All model training and testing are performed on NVIDIA A40.

\begin{table}[h]
\vspace{-0.2cm}
\caption{Performance comparison of different models on Youshu, NetEase and iFashion}
\vspace{-0.3cm}
\label{DBR:performance comparison}
\centering
\resizebox{0.79\textwidth}{!}{
\begin{tabular}{ccccccccccccc}
\toprule
\multirow{3}{*}{\textbf{Model}} & \multicolumn{4}{c}{\textbf{Youshu}}                                                           & \multicolumn{4}{c}{\textbf{NetEase}}                                                          & \multicolumn{4}{c}{\textbf{iFashion}}                                                         \\ \cline{2-13} 
                                & \multicolumn{2}{c}{{\ul \textbf{Metrics@20}}} & \multicolumn{2}{c}{{\ul \textbf{Metrics@40}}} & \multicolumn{2}{c}{{\ul \textbf{Metrics@20}}} & \multicolumn{2}{c}{{\ul \textbf{Metrics@40}}} & \multicolumn{2}{c}{{\ul \textbf{Metrics@20}}} & \multicolumn{2}{c}{{\ul \textbf{Metrics@40}}} \\
                                & \textbf{R}            & \textbf{N}            & \textbf{R}            & \textbf{N}            & \textbf{R}            & \textbf{N}            & \textbf{R}            & \textbf{N}            & \textbf{R}            & \textbf{N}            & \textbf{R}            & \textbf{N}            \\ \hline
\textbf{CrossCBR}               & 0.2786                & 0.1660                & 0.3794                & 0.1938                & 0.0795                & 0.0431                & 0.1204                & 0.0539                & 0.1146                & 0.0882                & 0.1649                & 0.1059                \\
\textbf{MultiCBR}               & 0.2699& 0.1572& 0.3667& 0.1833& 0.0906& 0.0492& 0.1357& 0.0610& 0.1495                & 0.1203& 0.2030& 0.1391\\
\textbf{BundleGT}               & 0.2907& 0.1720& 0.3951& 0.2006& 0.0891& 0.0472& 0.1366& 0.0597& 0.1222& 0.0935& 0.1774& 0.1129\\
\textbf{HyperMBR}               & 0.2710                & 0.1596                & 0.3730                & 0.1872                & 0.0727                & 0.0384                & 0.1146                & 0.0495& 0.0871& 0.0618& 0.1342& 0.0784\\
\textbf{MIDGN}                  & 0.2613                & 0.1492                & 0.3585                & 0.1756                & 0.0652& 0.0338& 0.1036& 0.0439& 0.0698& 0.0502& 0.1094& 0.0641\\
\textbf{BGCN}                   & 0.2390& 0.1392& 0.3425& 0.1673& 0.0657                & 0.0344                & 0.1052                & 0.0448                & 0.0626& 0.0443& 0.0990& 0.0571\\ \hline
\textbf{LightGCN}               & 0.2442                & 0.1422                & 0.3444                & 0.1696                & 0.0706                & 0.0375                & 0.1113                & 0.0483                & 0.0832                & 0.061                 & 0.1265                & 0.0762                \\
\textbf{NGCN}                   & 0.2361                & 0.1385                & 0.3326                & 0.165                 & 0.0601                & 0.0316                & 0.0985                & 0.0417                & 0.0675                & 0.048                 & 0.1067                & 0.0618                \\
\textbf{ItemKNN}                & 0.1942                & 0.1182                & 0.2578                & 0.1355                & 0.0206                & 0.0112                & 0.0332                & 0.0146                & 0.0791                & 0.0614                & 0.1129                & 0.0734                \\
\textbf{MFBPR}                  & 0.2543                & 0.1499                & 0.352                 & 0.1764                & 0.0647                & 0.0461                & 0.1018                & 0.0591                & 0.0647                & 0.0461                & 0.1018                & 0.0591                \\
\textbf{Pop}                    & 0.2511                & 0.1487                & 0.2896                & 0.1604                & 0.0308                & 0.0154                & 0.0502                & 0.0204                & 0.0221                & 0.0153                & 0.0384                & 0.021                 \\
\textbf{Random}                 & 0.0054                & 0.0026                & 0.0099                & 0.0039                & 0.0007                & 0.0003                & 0.0014                & 0.0005                & 0.0007                & 0.0004                & 0.0015                & 0.0007                \\ \bottomrule
\end{tabular}
}
\vspace{-0.3cm}
\end{table}

\textbf{Cooperative learning across bundle view and item view is effective.} We can observe that CrossCBR, MultiCBR and MIDGN which employ contrastive learning, and HyperMBR which leverages mutual learning, outperform BGCN which has no knowledge transfer between views. This demonstrates that cooperative learning strategy across multiple views, represented by contrastive learning and knowledge distillation, can align and contrast representations, as well as transfer learned knowledge to enhance representation quality, finally improve performance.

\textbf{Graph-based models have powerful high-order information encoding ability.} Graph models (BGCN, LightGCN, NGCF) are effective in bundle recommendation, which can be proved by their better performance that MFBPR. We attribute this to their superiority in capturing graph structure and multi-hop collaborative information.

\subsubsection{Generative BR}
\paragraph{\textbf{Experimental Details.}}
In the TripRec task, real-world POI check-in datasets from Flickr and Foursquare are often widely used for experimental evaluations \cite{gao2022selfsupervised, RANDOM, zhao2021photo2trip}, thus we utilize these for experiments. Eight cities are selected from the datasets, i.e., Edinburgh, Glasgow, Osaka, Toronto, Budapest, Vienna, Melbourne, and Tokyo. The precision, recall, and F1 score are commonly used in state-of-the-art TripRec methods \cite{wang2023query2trip, lim2015personalized, zhao2021photo2trip}. Considering that F1 score is calculated based on precision and recall, therefore the well-established metric of F1 score is chosen to evaluate performance.
We choose and reproduce 12 models: (i) Traditional TripRec: RANDOM \cite{RANDOM}, PersTour \cite{lim2015personalized}, RANK+Markov \cite{RANDOM}, POIRANK \cite{RANDOM}, and TRAR \cite{TRAR}. (ii) Neural TripRec: C-ILP \cite{he2019joint}, C-ALNS \cite{he2019joint}, CATHI \cite{CATHI}, DeepTrip \cite{DeepTrip}, PT-VPMF \cite{zhao2021photo2trip}, CTLTR \cite{CTLTR}, and Query2Trip \cite{wang2023query2trip}. The hyperparameter settings of our model are as follows: we set POI/trajectory embedding dimension as 256, batch size as 16, time embedding dimension as 32, hidden size as 128, learning rate as 0.5 (Trm), 0.01 (DAL), and 0.001 (DCL) respectively, the number of encoder layer as 5, and the number of attention heads as 8. We run model five times to average the results for experimental analysis.

\begin{table}[h]
\vspace{-0.3cm}
\caption{Performance comparison of different TripRec models in terms of F1}
\vspace{-0.3cm}
\label{GBR:performance comparison}
\centering
\resizebox{0.95\textwidth}{!}{
\begin{tabular}{ccccccccccccc}
\toprule
\multirow{2}{*}{\textbf{Datasets}} & \multicolumn{12}{c}{\textbf{Models}}                                                                                                                                                                                            \\
                                   & \textbf{RANDOM} & \textbf{PersTour} & \textbf{POIRANK} & \textbf{RANK+Markov} & \textbf{C-ILP} & \textbf{C-ALNS} & \textbf{CATHI} & \textbf{TRAR} & \textbf{DeepTrip} & \textbf{PT-VPMF} & \textbf{CTLTR} & \textbf{Query2Trip} \\ \hline
\textbf{Edi.}                      & 0.570           & 0.656             & 0.700            & 0.659                & 0.769          & 0.768           & 0.772          & 0.770         & 0.765             & 0.710            & 0.853          & 0.858               \\
\textbf{Gla.}                      & 0.632           & 0.801             & 0.768            & 0.754                & 0.853          & 0.852           & 0.815          & 0.820         & 0.831             & 0.817            & 0.874          & 0.890               \\
\textbf{Osa.}                      & 0.621           & 0.686             & 0.745            & 0.715                & 0.763          & 0.762           & 0.758          & 0.770         & 0.834             & 0.809            & 0.889          & 0.895               \\
\textbf{Tor.}                      & 0.621           & 0.720             & 0.754            & 0.723                & 0.818          & 0.815           & 0.807          & 0.801         & 0.808             & 0.779            & 0.874          & 0.884               \\
\textbf{Bud.}                      & 0.289           & 0.748             & 0.784            & 0.752                & 0.832          & 0.829           & 0.819          & 0.820         & 0.837             & 0.823            & 0.842          & 0.852               \\
\textbf{Vie.}                      & 0.351           & 0.651             & 0.662            & 0.645                & 0.750          & 0.747           & 0.737          & 0.728         & 0.730             & 0.719            & 0.785          & 0.803               \\
\textbf{Mel.}                      & 0.558           & 0.483             & 0.637            & 0.613                & 0.729          & 0.728           & 0.723          & 0.713         & 0.683             & 0.677            & 0.778          & 0.793               \\
\textbf{Tok.}                      & 0.581           & 0.701             & 0.728            & 0.677                & 0.835          & 0.833           & 0.809          & 0.812         & 0.826             & 0.820            & 0.860          & 0.878               \\ \bottomrule
\end{tabular}
}
\vspace{-0.3cm}
\end{table}

\textbf{Performance of seq2seq-Based Neural TripRec.} Generally, seq2seq model-based neural TripRec methods (CATHI, DeepTrip, CTLTR, and Query2Trip) yield better results than other TripRec methods. This confirms that the seq2seq model as the basic framework is effective. Among the compared models (CATHI, DeepTrip, and CTLTR), Query2Trip outperforms CTLTR by 1.47\% (F1) on average. There are two possible reasons. The first reason is that Query2Trip treats the user’s given query as an explicit preference signal to learn debiased representations from positives and negatives to mitigate selection bias and exposure bias. Besides, its Transformer with position coding architecture ensures the global coherence of the query sequence and trajectory sequence.

\textbf{Performance of Non-seq2seq-Based Neural TripRec.} Even though CILP, C-ALNS, and PT-VPMF are deep learning-based methods, seq2seq-based models are not adopted. Furthermore, they regard the user-specific query as a post-processing constraint and ignore the latent preference information contained in the query. Query2Trip method exhibits an advantage over PT-VPMF by an obvious margin, where the improvement achieves 11.66\% (F1) on average. Also, it can be seen that basic embedding is far from enough for performance improvement. This drives researchers to design dual-debiased learning to improve representation ability.

\section{Challenges and Trends}
\label{section6:challenges and trends}
While existing works have established a solid foundation for bundle recommendation research, further opportunities remain. In this section, we discuss the main challenges and future trends.

\subsection{Representation Learning in Non-Euclidean Spaces}
Currently, most existing bundle recommendation work focuses on learning representations in Euclidean space \cite{AttList, BGCN, GRAM-SMOT, nguyen2024bundle, sang2024multi}. Euclidean space is argued more suitable for handling grid-like structures, and produce high distortion when dealing with data that have complex structures \cite{chami2019hyperbolic}. Bourgain theory \cite{linial1995geometry} and other works \cite{zhang2021where, sala2018representation} have demonstrated that Euclidean space struggles to embed tree-like structured data with low distortion. Therefore, exploring representation learning in non-Euclidean spaces is not only challenging but also a very attractive research direction. In the field of bundle recommendation, interaction data often exhibit an underlying tree-like structure, which is more suitable for learning representations in the hyperbolic space with negative curvature. Some studies have began to attempt bundle recommendation research in hyperbolic space \cite{HBGCN, HyperMBR}. These methods map Euclidean embeddings to hyperbolic space. \textbf{\textit{However, since there is no corresponding graph convolution operation in hyperbolic space, to perform equivalent graph convolution operations as in Euclidean space, one needs to map the data to the tangent space, complete the graph convolution operations, and then map the results back to hyperbolic space. This mapping and inverse mapping process introduces unnecessary noise, resulting the deduction of the accuracy of the embeddings.}} In future work, finding a computational method that allows graph convolution operations to be performed entirely within hyperbolic space without relying on the tangent space could further improve the performance.

Additionally, studies \cite{sun2024contrastive, sun2023self} have began to explore the theories and methods of graph learning in Riemannian space, a type of non-Euclidean space that can encompass different curvatures (positive, zero, and negative curvatures). CSincere \cite{sun2024contrastive} discovered that the users and items differ in terms of both $\delta$-hyperbolicity and degree distribution, and believes that it is more rational to model the users and items in two different spaces. Riemannian geometry provides the notion of curvature to distinguish the structure pattern between different spaces. In bundle recommendation, the degree distributions of users, bundles, and items are different, and the $\delta$ values of the three types of interaction graphs are not the same. Therefore, modeling different entities in different spaces is reasonable and feasible. In addition, previous studies, whether in Euclidean or hyperbolic space, have tried to learn embeddings in a fixed curvature space. However, user interactions will continue to generate, requiring a dynamically changing curvature space to represent learning for users, items, and bundles. In summary, future research directions can be explored in two aspects: First, \textbf{\textit{researchers can enrich the content of the datasets, supplement information for the existing mainstream datasets used in bundle recommendation research}}, such as adding community attributes and temporal dynamics, or build such datasets to lay the foundation for subsequent representation learning in adaptive Riemannian space. Second, \textbf{\textit{researchers can consider using adaptive Riemannian space for representation learning}}. Moreover, considering that the curvature of the graph may change over time, incorporating the variation of graph curvature into the bundle recommendation model can make it more flexible to adapt to environmental changes, improving the model's adaptability and accuracy.

\subsection{Data Sparsity and Cold-Start Issues of Discriminative BR}
\label{data sparsity and cold start}
\textbf{\textit{Data sparsity and cold-start problems remain major challenges.}} The problem of skewed distribution of bundle interaction is prevalent by analyzing three mainstream datasets \cite{CoHEAT, li2025divide}, i.e., a small number of bundles get the vast majority of interactions, while a large number of bundles have little or no interactions at all. When new bundles are added to the system, it is difficult for the system to recommend them effectively due to the lack of historical interaction data. Recent research \cite{CoHEAT} has proposed a solution for cold-start bundle recommendation, which solves the problem of highly skewed distribution of bundle interactions by taking into account the popularity of the bundles, and relying more on the user-item view, which provides richer information, for less popular bundles. \textbf{\textit{To alleviate the data sparsity and cold-start problem in bundle recommendation, a promising future direction is to introduce multimodal information (e.g., images, descriptions of items or bundles, etc. ) to model user preferences harnessing LLMs.}} There have been many studies \cite{ma2024cirp, liu2024harnessing, kim2024less, ma2024leveraging} that have began to introduce multimodal information and attempt to utilize LLMs to boost recommendation performance.

\subsection{User Intents and Dynamic Bundle Generation of Generative BR}
Early bundle generation methods usually relied on specifying hard constraints to create bundles \cite{zhu2014bundle, beladev2016recommender, garfinkel2006design, xie2010breaking, xie2014generating}. With the advancement of technology, deep learning-based approaches \cite{BGN, BGGN, wei2022towards} have gradually been introduced to capture latent associations between items for bundle generation. However, these methods are often limited to creating fixed-size bundles or constrained to generating a single bundle for each user. More importantly, \textbf{\textit{they neglect the underlying user intents reflected by the bundles in the generation process, resulting in less intelligible bundles.}} Actually, items in a high-quality bundle should all reflect the same user intent, describing a consistent user need. For instance, if a user is shopping a camera, identifying the \textquotesingle photography\textquotesingle\ intent would allow bundle recommendation system to generate a photography kit consisting of lenses, a tripod, and memory cards.

There have been studies \cite{sun2024adaptive, sun2024revisiting, li2023next} focusing on exploring the bundle generation that reflect user intents. The remarkable advancement of Large Language Models (LLMs) has led to widespread adoption for more effective recommendation \cite{harte2023leveraging, dai2023uncovering, he2023large}. Notably, several studies \cite{dai2023uncovering, he2023large} highlight ChatGPT's potential to mitigate the cold start issue and provide explainable recommendations. Furthermore, studies \cite{sun2024adaptive} leverage the advanced reasoning capability of LLMs to infer user intent from user sessions through adaptive contextual learning paradigm and generate personalized bundles based on these intents. Inferring user intents not only helps to generate bundles that better meet the actual needs of users but is also key to enhancing the transparency of recommendation systems and user trust. \textbf{\textit{Future directions can further explore how to more accurately capture and interpret user intents.}} On one hand, researchers can investigate how to design better contextual examples and prompts to enhance the interpretability of LLMs. On the other hand, future work could consider combining various types of data such as images and videos to provide richer explanations and recommendations. \textbf{\textit{Moreover, considering that user preferences and intents change over time, the dynamic capture of user intent and the implementation of dynamic bundle generation become particularly important.}} The work \cite{yang2024nonautoregressive} proposes plans to explore dynamic bundle generation methods. \textbf{\textit{Therefore, future research trends can focus on developing dynamic bundle generation methods that adapt to changes in user intent, and how to provide more interpretable recommendations through these methods.}}

\subsection{Responsible Bundle Recommender System}
\textbf{\textit{Existing recommender systems typically focus on accuracy and personalization but increasingly call for an effective means of ensuring the systems work responsibly.}} Without appropriate responsible technologies, recommender systems can have unintended impacts on users, communities, and society. For instance, recommendation algorithms trained on imbalanced data may be biased toward catering to the preferences of the majority, neglecting minority groups; systems without measures against misinformation can amplify its spread.

Since a bundle is composed of several items, the bias issues in item recommendation can amplify issues such as bias and unfairness in bundle recommendation. We conduct an empirical study to analyze popularity bias on three datasets (Youshu, NetEase, iFashion). We find that the distribution of interactions is noticeably imbalanced, with some items receiving inadequate exposure, which exacerbates the unfairness among items and suggests the presence of bias in bundle recommendation. In addition, as discussed in the Section \ref{data sparsity and cold start}, these mainstream datasets generally suffer from a skewed distribution of user-bundle interactions, leading to a long-tail phenomenon where popular or frequently interacted bundles are recommended while neglecting less popular bundles, even though they better meet user needs. This popularity bias reduces fairness in recommendations and further marginalizes less popular bundles, exacerbating data imbalance.

To address the popularity bias issue in bundle recommendation, Jeon et al. \cite{PopCon} introduces a debiasing method that reduces bias through popularity-based negative sampling and enhances diversity with a reranking algorithm. For less popular bundles, Jeon et al. \cite{CoHEAT} dynamically adjusts view weights and uses curriculum learning to focus on more informative affiliation view. Meanwhile, Li et al. \cite{li2024boosting} noticed that the spread of misinformation may be amplified and proposed a healthiness-aware and category-wise meal recommendation model, which boosts healthiness exposure by using nutritional standards as knowledge to guide the model training. However, these methods do not fully solve the problem, and responsible bundle recommender systems remain an extremely necessary and challenging direction. Coppolillo et al. \cite{Coppolillo2024balanced} introduced a quality measure that rewards debiasing techniques that successfully push a recommender system to recommend niche items without sacrificing global recommendation accuracy. Future research can investigate whether new mitigation strategies in bundle recommendation can be defined with the related quality measures. Moreover, in the real world, biases are usually dynamic rather than static. For example, the fashionability of outfit changes frequently; the composition of items within bundles is updated; users interact with new bundles daily. In summary, biases tend to evolve over time. \textbf{\textit{Exploring how biases evolve and analyzing how dynamic biases affect recommendation systems will be an important research direction in responsible bundle recommendation.}}

\subsection{Large Language Models for Bundle Recommendation}
In recommendation system, there has been a clear shift toward generative paradigms, where retrieval and ranking are reformulated as sequence modeling tasks \cite{Rajput2023recommender, zhai2024actions}. These advances highlight the potential of generative modeling to improve scalability, alleviate cold-start issues, and better capture sequential dependencies. Inspired by this trajectory, researchers have begun to investigate how such ideas can be tailored to bundle recommendation. Within bundle recommendation, a first line of work has leveraged multimodal foundation models to capture richer semantic information for bundle construction \cite{ma2024leveraging, RaMen, ma2024cirp}. By integrating features from text, images, audio, and user interaction data, these approaches uncover collaborative relations among items that extend beyond co-occurrence statistics. For example, methods such as RaMen \cite{RaMen} and CLHE \cite{ma2024leveraging} employ multimodal encoders and contrastive learning to unify heterogeneous signals, thereby enhancing bundle representation learning and alleviating data sparsity. This demonstrates the importance of multimodal feature extraction for modeling the complex structures inherent in real-world bundles.

Building on these foundations, large language models (LLMs) have recently emerged as powerful tools for bundle recommendation, owing to their broad knowledge and strong reasoning capabilities \cite{liao2024LLaRA, zheng2024adapting, zhang2025collm}. Existing efforts can be broadly grouped into two directions. The first focuses on improving item and bundle representations \cite{Bayrak2025enhancing, liu2025finetuning}. For instance, \cite{Bayrak2025enhancing} uses LLMs for offline data augmentation, enriching interaction data for embedding training. Similarly, Bundle-MLLM \cite{liu2025finetuning} unifies textual, visual, relational, and behavioral information into a tokenized input, allowing LLMs to form coherent multimodal representations through attention-based fusion. The second direction treats LLMs as generative engines for bundle construction \cite{liu2025finetuning, sun2024adaptive, zhu2023text2bundle, feng2025routing, feng2025doesknowledge}. Representative examples include AICL \cite{sun2024adaptive}, which adapts LLMs with retrieval-augmented demonstrations and self-correction; Text2Bundle \cite{zhu2023text2bundle}, which combines intent extraction with reinforcement learning for bundle generation; and RouteDK \cite{feng2025routing}, which employs knowledge distillation and mixture-of-experts to balance efficiency and knowledge consistency. Knowledge distillation has further been explored to reduce costs without degrading bundle generation quality \cite{feng2025doesknowledge}. These studies provide a more comprehensive and forward-looking perspective for LLMs in bundle recommendation.

Looking ahead, promising directions include positioning LLMs directly as bundle recommenders - either serving as semantic rankers over candidate bundles or evolving toward end-to-end generative recommenders such as OneRec \cite{deng2025onerec}. Nevertheless, several key challenges remain. First, scaling LLMs to \textit{\textbf{handle the long and complex sequences involved in candidate bundle sets}} is critical. Second, stronger \textit{\textbf{controllability over output formats}} is needed to ensure generated bundles meet system requirements. Third, reliability must be addressed through \textit{\textbf{self-correction and feedback mechanisms}} to mitigate hallucinations during inference. Finally, \textit{\textbf{more effective integration of multimodal content information \cite{sheng2025language, zheng2024adapting}}}—including textual, visual, and behavioral data — is required to enable deeper personalization and stronger user-centric bundle recommendation.

\section{Conclusions}
\label{section7:conclusion}
Bundle recommendation, as an emerging and evolving research direction, has gained wide attention in various application scenarios. With the development of new techniques such as graph learning, knowledge distillation and so on, the research on bundle recommendation has made further progress in recent years. A comprehensive review on bundle recommendation is necessary and meaningful. Our survey introduces a taxonomy that categorizes it into two lines: discriminative bundle recommendation and generative bundle recommendation. We systematically summarize and review representation learning, prediction interaction, and bundle generation methods within these two lines. Meanwhile, we provide a thorough summary of existing resources for bundle recommendation (including datasets and evaluation metrics), and use these resources to reproduce and compare the performance of mainstream models. Finally, we discuss the main challenges and potential future research directions. This survey aims to help readers get a quick overview of the developments and key aspects of bundle recommendation and inspire future research.

\begin{acks}
This work is partially supported by NSFC, China (No.62276196), the research grant (RGPIN-2020-07157) from the Natural Science and Engineering Research Council (NSERC) of Canada and York Research Chairs (YRC) program. The authors also gratefully appreciate the anonymous reviewers and associate editor for their valuable comments and constructive suggestions that greatly helped to improve the quality of the paper.
\end{acks}


\bibliographystyle{unsrt}

\def\bibfont{\fontsize{5.26}{6.9}\selectfont}
\bibliography{bundleRS_survey}


\appendix
\end{document}